\documentclass[%
 reprint,
showpacs,preprintnumbers,
 amsmath,amssymb,
 aps,
 prl,
]{revtex4-1}

\usepackage{graphicx}
\usepackage{dcolumn}
\usepackage{bm}
\usepackage{hyperref}
\usepackage[mathlines]{lineno}
\usepackage{epstopdf}
\usepackage{float}
\usepackage{siunitx}
\usepackage[colorinlistoftodos,prependcaption,textsize=tiny]{todonotes}
\DeclareSIUnit\gauss{G}
\renewcommand\vec{\mathbf}

\begin{document}


\title{Two-Dimensional Homogeneous Fermi Gases}

\author{Klaus Hueck}
\email{khueck@physik.uni-hamburg.de}
\author{Niclas Luick}
\author{Lennart Sobirey}
\author{Jonas Siegl}
\author{Thomas Lompe}
\author{Henning Moritz}
\affiliation{Institut f\"ur Laserphysik, Universit\"at Hamburg, Luruper Chaussee 149, 22761 Hamburg, Germany.}

\date{\today}

\begin{abstract}
We report on the experimental realization of homogeneous two-dimensional (2D) Fermi gases trapped in a box potential. 
In contrast to harmonically trapped gases, these homogeneous 2D systems are ideally suited to probe local as well as non-local properties of strongly interacting many-body systems.
As a first benchmark experiment, we use a local probe to measure the density of a non-interacting 2D Fermi gas as a function of chemical potential and find excellent agreement with the corresponding equation of state (EOS).
We then perform matter wave focusing to extract the momentum distribution of the system and directly observe Pauli blocking in a near unity occupation of momentum states. 
Finally, we measure the momentum distribution of an interacting homogeneous 2D gas in the crossover between attractively interacting fermions and bosonic dimers.
\end{abstract}

\maketitle

Ultracold 2D Fermi gases are uniquely suited to investigate the interplay of reduced dimensionality and strong interactions in quantum many-body systems in a clean and well-controlled environment.
Experiments have reported on the creation of 2D Fermi gases with equal \cite{Martiyanov10,Dyke11} and unequal spin populations \cite{Ong15,Mitra16} and investigated pairing \cite{Frohlich11,Feld11,Sommer12,Cheng16}, Fermi-liquid \cite{Frohlich12} and polaron physics \cite{Koschorreck12,Zhang12}. 
The EOS \cite{Makhalov14,Fenech16,Boettcher16} was determined and evidence for pair condensation \cite{Ries15} and for a Berezinskii-Kosterlitz-Thouless (BKT) transition \cite{Murthy15} could be observed. 
Yet so far, ultracold 2D Fermi gases have always been studied in harmonic trapping potentials, which qualitatively change the density of states and give rise to inhomogeneous density distributions.
This hinders the observation of critical phenomena with diverging correlation length and exotic phases such as the Fulde-Ferrell-Larkin-Ovchinnikov (FFLO) state \cite{Fulde64,Larkin65,Conduit08,Toniolo17}.
Furthermore, the inhomogeneous density distribution complicates the interpretation of non-local quantities such as correlation functions or momentum distributions, which can only be extracted as trap-averaged quantities \cite{Ries15,Murthy15}. 

These issues can be overcome by creating homogeneous gases in box potentials whose walls are formed by repulsive optical dipole potentials. 
Following this method, three-dimensional (3D) uniform Bose gases have recently been realized and used to investigate coherence and thermodynamic properties \cite{Gaunt13,Schmidutz14} as well as non-equilibrium dynamics \cite{Navon15}.
In homogeneous 2D Bose gases, the emergence of condensation, vortices and supercurrents were studied \cite{Corman14,Chomaz15}.
Very recently, the creation of 3D Fermi gases in a box potential has been demonstrated, Pauli blocking in momentum space was observed and both balanced and imbalanced Fermi gases have been studied in the strongly-interacting regime \cite{Mukherjee17}.

Here, we report on the experimental realization of homogeneous 2D Fermi gases with tunable interactions. 
By preparing a non-interacting Fermi gas we realize a textbook example of statistical physics and directly observe Pauli blocking in the occupation of momentum states. 

\begin{figure}
	\center
	\includegraphics[width = \linewidth]{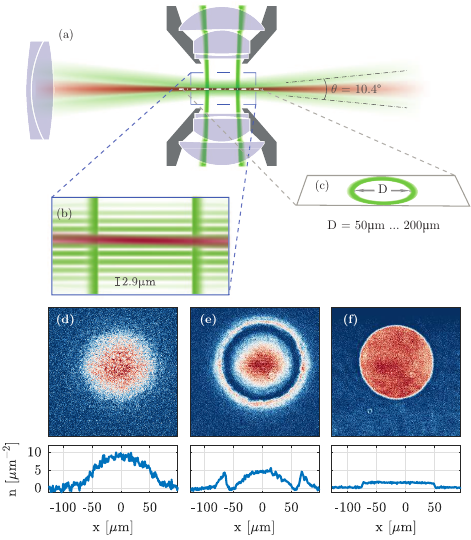}
	\caption{\label{fig:Experiment}Sketch of the experimental setup: 
	The atoms are loaded from a highly elliptic red detuned optical trap (red) into a single nodal plane of a blue detuned optical lattice (light green) which is formed by two laser beams ($\lambda = \SI{532}{\nano\meter}$) intersecting under an opening angle of $\theta = \SI{10.4}{\degree}$ (a,b).
	The radial confinement is provided by a ring-shaped repulsive potential (dark green) whose diameter D can be adjusted between $\SI{50}{\micro\meter}$ and  $\SI{200}{\micro\meter}$ (c).
	Panels (d-f) show averaged (20 - 50 images) in situ density profiles and respective central line cuts at different stages of the preparation of a strongly interacting homogeneous Fermi gas at $B = \SI{830}{\gauss}$:  
	After evaporation in the elliptic trap (d), the outer, high entropy region of the cloud is cut away by the repulsive ring potential (e). 
	After further evaporation, the radial magnetic confinement is ramped down to spill the atoms outside the ring and the atoms are transferred into the lattice and we obtain a homogeneous 2D gas (f).}
\end{figure}

\noindent To measure the momentum distribution of interacting gases we have established a technique to rapidly remove one spin component and thereby project the system onto a non-interacting state.
We apply this technique to a gas with intermediate attractive interactions and observe a momentum distribution that is qualitatively similar to that of a non-interacting gas.


We perform our experiments with an equal spin mixture of $^6$Li atoms in the lowest two hyperfine states $|F,m_F\rangle = |\frac{1}{2},\frac{1}{2}\rangle$ and  $|\frac{1}{2},-\frac{1}{2}\rangle$, which we designate as $\left|\uparrow\right>$ and $\left|\downarrow\right>$, respectively. 
A sketch of the experimental setup is shown in Fig.~\ref{fig:Experiment}(a).
The atoms are pre-cooled as described in \cite{Weimer15} and then transferred into a hybrid trap consisting of a highly elliptic red detuned optical trap and a variable radial magnetic confinement, which is generated by the curvature of the magnetic offset field used to tune the interparticle interactions \cite{Supp}.
This variable trapping can be used to compensate for anti-confinement introduced by the lattice potential, which provides the 2D confinement described below. 
After forced evaporative cooling in the elliptic trap (Fig.~\ref{fig:Experiment}(d)), we ramp on a repulsive optical ring potential as sketched in Fig.~\ref{fig:Experiment}(b,c).
This ring potential is generated by a cascaded setup of three axicons and projected onto the atoms using a high-resolution (NA = 0.62) objective \cite{McLeod54,Manek98,Supp,Mukherjee17}. 
We use the ring to cut out the central, low-entropy part of the cloud (Fig.~\ref{fig:Experiment}(e)) and then ramp down the radial magnetic confinement such that the excess atoms outside the ring leave the observation volume.

Next, we bring the gas into the 2D regime by loading it into a blue detuned optical lattice in $z$-direction.
In this lattice, the level spacing $\hbar\omega_z = h \cdot\SI{12.4\pm0.1}{\kilo\hertz}$ between the ground and first excited state in the vertical direction exceeds both the highest chemical potential $\mu < h\cdot\SI{4}{\kilo\hertz} $ and the highest thermal energy $k_B T < h\cdot \SI{2}{\kilo\hertz}$ encountered during our experiments and hence the system is in the 2D regime \cite{Petrov00,Dyke16}.

To transfer the atoms into a single node of the lattice, we recompress the cloud by ramping up the power of the elliptic trap, which reduces the width of the cloud in $z$-direction below the lattice spacing of \SI{2.9}{\micro\meter}. 
By optimizing the position of the elliptic trap with respect to the lattice, optimally \SI{93}{\percent} of the atoms can be loaded into a single layer, where the number of atoms in adjacent layers can be determined to high precision in a single shot matter wave focusing measurement \cite{Supp}. 
By shifting the $z$-position of the elliptic trap by half a lattice period, it is also possible to create two equally populated adjacent layers \cite{Hadzibabic06,Supp}.
This makes the loading of non-interacting gases more robust against populating adjacent layers by thermally excited atoms and furthermore doubles the recorded signal for absorption imaging. 

\begin{figure}
	\center
	\includegraphics[width = \linewidth]{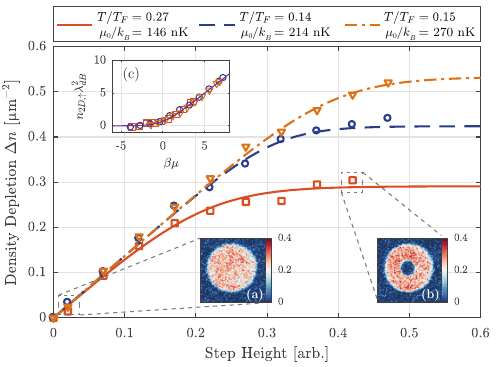}
	\caption{\label{fig:2DEOS}
	Density EOS for non-interacting homogeneous 2D Fermi gases:
	The EOS is mapped out for different densities and temperatures by imprinting a repulsive potential step onto the atoms.
	This causes a density depletion $\Delta n$ in the center of the cloud (a,b). 
	Measuring this density depletion $\Delta n$ as a function of the step height directly yields the density EOS of the system.
	By fitting the data with the EOS of a non-interacting Fermi gas we extract the temperature $T$ and chemical potential $\mu_0$ for each dataset.
	The higher $T/T_F$  for the dataset having the lowest density in the outer ring (red squares) is most likely due to a reduced evaporation efficiency.
	Using the fit results for $T$ and $\mu$ to rescale the data and plotting the dimensionless quantity $n_{\text{2D},\uparrow}\lambda_{dB}^2$ causes the different datasets to collapse onto a single curve (c). 
	The data shows excellent agreement with the prediction for a non-interacting 2D Fermi gas (solid purple line).}
\end{figure}

In a first series of experiments, we study a non-interacting Fermi gas, which provides us with a well-defined starting point for our exploration of interacting systems.
To create such non-interacting systems we first prepare a dual layer homogeneous 2D Fermi gas at a magnetic offset field of $B=\SI{320}{\gauss}$.
At this field the gas is weakly interacting with a 3D scattering length of $a_{\text{3D}} = -290\,a_0$, where $a_0$ is the Bohr radius.
We perform further evaporative cooling by slowly decreasing the height of the confining ring potential and then ramp to $B=\SI{527}{\gauss}$, which is close to the zero crossing of the scattering length, to obtain a non-interacting Fermi gas. 

As a first benchmark experiment, we measure the density EOS $n_{\text{2D},\uparrow}(\mu, T)$ of this non-interacting Fermi gas.
We imprint a potential step, which is generated by a blue detuned laser beam reflected off the surface of a digital micromirror device (DMD) and projected onto the atoms \cite{Hueck17}. We then image the resulting density distribution using high-intensity absorption imaging \cite{Reinaudi07,Hueck171,Supp}.
As shown in Fig. \ref{fig:2DEOS}(a,b), the repulsive potential causes a disk shaped density depletion in the center of the cloud which covers about \SI{10}{\percent} of its area.
We apply potential steps with different heights $V$ while observing the corresponding density depletion $\Delta n(V) = n_{\text{2D},\uparrow}^{\text{disk}}-n_{\text{2D},\uparrow}^{\text{center}}(V)$, where $n_{\text{2D},\uparrow}^{\text{disk}}$ and $n_{\text{2D},\uparrow}^{\text{center}}$ correspond to the single layer density in the undisturbed and depleted parts of the trap, respectively.
We perform such EOS measurements for gases with different densities and temperatures; the resulting datasets are shown in Fig. \ref{fig:2DEOS}.

We calibrate the potential step height $V$ by performing a linear Thomas-Fermi fit to the first four points of the different EOS measurements and take the mean of the resulting values \cite{Boettcher16}.
To extract the temperature and chemical potential, we fit the density depletion with $\Delta n(\mu_0, T, V)=n_{\text{2D},\uparrow}(\mu_0,T)-n_{\text{2D},\uparrow}(\mu_0-V,T)$ using the theoretical EOS $n_{\text{2D},\uparrow}(\mu,T)=\lambda_{dB}^{-2}\log[1+\exp(\mu\beta)]$ for a non-interacting 2D Fermi gas \cite{Bauer14}.
Here, $\beta = (k_BT)^{-1}$ and the thermal de Broglie wavelength is given by $\lambda_{dB} = \sqrt{2\pi\hbar^2/mk_BT}$, where $m$ is the mass of a $^6$Li atom.
We approximate the chemical potential $\mu_0$ in the outer part of the trap to be constant for all step heights.
For our coldest dataset we obtain a temperature of $T/T_F = \num{0.14\pm0.02}$, where the Fermi temperature $T_F$ is calculated from $T$ and $\mu_0$ using $T_F=T\log[1+\exp(\beta\mu_0)]$ \footnote{This equation is found by solving the 2D EOS for $\mu_0$, taking the $T=0$ limit, yielding $T_F=n_{\text{2D},\uparrow}\,2\pi\hbar^2/(m k_B) $ and reinserting the EOS for $n_{\text{2D},\uparrow}$}.

We validate these measurements by plotting the dimensionless quantity $n_{\text{2D},\uparrow}\lambda_{dB}^2$ as a function of $\beta \mu$ for each of the different systems (Fig.~\ref{fig:2DEOS}(c)) \cite{Hung11}.
The different datasets all collapse onto a single curve and are in excellent agreement with the theoretical expectation. 

\begin{figure}
	\center
	\includegraphics[width = \linewidth]{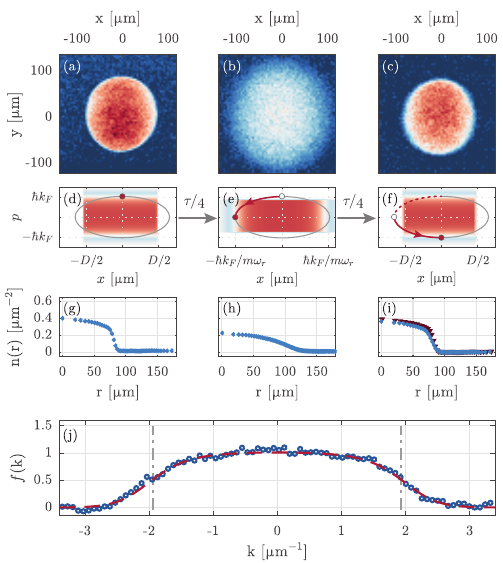}
	\caption{\label{fig:FermiGas} Momentum distribution of a non-interacting 2D Fermi gas:
	To measure the momentum distribution, we switch off the confining ring potential and let the gas evolve in a weak harmonic potential.
	A free time evolution $t$ for a quarter of the trap period $\tau$ performs a rotation in phase space by \SI{90}{\degree} as sketched in (d,e), causing the momentum distribution of the gas to be mapped into real space.
	Averaged images (51 - 62 realizations) and corresponding azimuthal averages of the density and momentum distribution are shown in (a,b) and (g,h) respectively.
	After a free time evolution of half a trap period, the in situ density distribution is mapped back to real space (c);
	the azimuthal averages at $t = 0$ (red triangles) and $t = \tau/2$ (blue dots) are almost identical (i). 
	A diagonal cut through the momentum distribution (b) reveals the occupation $f(k)$ of the system (j), which shows close to unity occupation around $k=0$ and drops off at the Fermi wave vector $k_F =\SI{1.93\pm0.02}{\micro\meter^{-1}}$ (gray dash dotted lines). 
	A fit with a Fermi distribution (red dashed line) yields a temperature of $T/T_F = \num{0.31\pm 0.02}$.}
\end{figure}
We now go beyond this local probing of density and chemical potential by performing a direct measurement of the momentum distribution of an ideal 2D Fermi gas. 
We achieve this by mapping the momentum distribution to real space using matter wave focusing \cite{Shvarchuck02,Tung10,Murthy14,Ries15}: 
We switch off the radial confinement provided by the ring potential and let the system evolve for a time $t$ in a weak harmonic potential in radial direction. 
After a time-evolution of a quarter of the radial trap period $\tau = 2\pi/\omega_r$, all particles with momentum $\hbar \vec k$ have moved to a position $\vec r = \hbar \vec k/m\omega_r$. 
Hence, the momentum distribution $\tilde{n}(\vec k)$ can be directly extracted from the density distribution $n(\vec r,t)$ at $t=\tau/4$ via $\tilde{n}(\vec k) = (\hbar/m\omega_r)^2 \cdot n(\vec r = \hbar \vec k/m\omega_r, \tau/4)$ (Fig.~\ref{fig:FermiGas}(b,e,h)).

This technique can also be extended to perform matter wave imaging instead of matter wave focusing by letting the system evolve for $t = \tau/2$ instead of $t = \tau/4$ \footnote{In our case this imaging has a magnification of 1. Other magnifications are accessible by switching to a different radial trap frequency after the $\tau/4$ point.}.
This causes the initial density distribution to reappear inverted around the center of the trap, i.e. $n(\vec r,\tau/2) = n(-\vec r,0)$.
Comparing the matter wave imaged distribution at $t = \tau/2$ with the initial distribution provides a measure for the quality of the matter wave lens formed by the radial potential, which can be affected by anharmonicities of the potential.
For our experiments, we set the radial magnetic trap frequency to a value of $\omega_r = 2\pi \cdot \SI{33.3\pm0.5}{\hertz}$ and ramp down the depth of the $z$-confinement by a factor of five to minimize the influence of its anti-trapping potential while keeping the atoms in the depth of field. 
We find that the in situ and matter wave imaged density distributions are virtually indistinguishable (Fig.~\ref{fig:FermiGas}(a,c,i)), which shows that for this non-interacting system, our matter wave focusing gives an accurate measurement of the momentum distribution.

To extract the occupation $f(\vec k) = A_k \cdot \tilde{n}(\vec k)$ from the momentum distribution $\tilde{n}(\vec k)$, we use the $k$-space area $A_k = 16 \pi / \text{D}^2 $ of a single $k$-mode in a box potential with diameter D.
This allows us to directly observe Pauli blocking in our non-interacting Fermi gas, which manifests itself in a unity occupation of $k$ modes around $k = 0$, followed by a drop in the occupation at the Fermi wave vector $k_F$ (Fig.~\ref{fig:FermiGas}(j)).

Next, we quantitatively determine the chemical potential and the temperature of the gas by fitting our data with the Fermi distribution
$$ f\left(k\right) = \frac{\zeta}{1+\exp\left[\beta\left(\frac{\hbar^2k^2}{2m}-\mu_0\right)\right]}.$$
The free parameters of the fit are the temperature $T$, the chemical potential $\mu_0$ and an overall amplitude $\zeta$ which accounts for systematic errors in the determination of $\tilde{n}(\vec k)$ and $A_k$.
The fit is in excellent agreement with the data (Fig.~\ref{fig:FermiGas}(j)) and yields a chemical potential $\mu_0 = k_B\SI{148.8\pm 2.6}{\nano\kelvin}$, a temperature $T = \SI{46.7\pm 2.2}{\nano\kelvin}$ and $\zeta = 1.05\pm0.06$, where the errors denote $1\sigma$-confidence intervals of the fit.
The dominant sources of systematic errors on the amplitude of $f(k)$ are the \SI{2}{\percent} uncertainty of the radial trap frequency $\omega_r$, the \SI{7}{\percent} uncertainty in the density calibration and the \SI{4}{\percent} uncertainty in the determination of the ring diameter D from the in situ images.
The fit results translate to $T/T_F = \num{0.31\pm 0.02}$, $\mu_0/\hbar \omega_z = 0.250\pm0.005$, and a Fermi wave vector $k_F =\SI{1.93\pm0.02}{\micro\meter^{-1}}$.
This is in very good agreement both with the Fermi wave vector deduced from the mean density $k_{F,\bar{n}} = \sqrt{4\pi \bar{n}_{\text{2D}, \uparrow}} = \SI{1.86\pm0.08}{\micro\meter^{-1}}$ and the temperature and chemical potential obtained for a similar evaporation depth in the EOS measurement shown in Fig.~\ref{fig:2DEOS} (red solid line) \footnote{We chose a low evaporation depth, despite the fact that we achieve our lowest $T/T_F$ at higher Fermi energies, since we want $k_F$ to be small enough that the full momentum distribution is captured by the field of view of the imaging system.}.
We note that the fitted temperature is an upper bound, affected by fluctuations in the particle number and the inhomogeneity of the density distribution, which is smaller than \SI{11}{\percent} of the mean density \cite{Supp}.
This value includes both the actual density inhomogeneity due to the presence of the harmonic potential used for the matter wave focusing and artifacts due to imperfections of the imaging beam.

\begin{figure}
	\center
	\includegraphics[width = \linewidth]{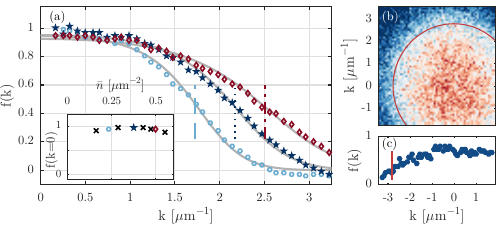}
	\caption{\label{fig:Fig4} Saturation in the occupation of momentum states. 
	The occupation $f(k)$ of non-interacting Fermi gases with in-situ densities of $\bar n = \SI{0.24}{\micro\meter^{-2}}$ (light blue hexagons), $\SI{0.38}{\micro\meter^{-2}}$ (dark blue stars) and $\SI{0.50}{\micro\meter^{-2}}$ (red diamonds) is shown in (a).
	For low momenta we find a near unity occupation that is independent of the in-situ density (see inset) which is direct evidence of Pauli blocking.
	The Fermi wave vectors deduced from Fermi fits to the distribution agree well with the $k_{F,\bar{n}}=\sqrt{4\pi \bar{n}_{\text{2D}, \uparrow}}$ (vertical lines) calculated from the in-situ density.
	An image of the momentum distribution of an attractively interacting Fermi gas is shown in panel (b) and a cut through the distribution in panel (c). 
	The Fermi momentum $k_{F,\bar{n}}$ for a non-interacting gas with equal density is indicated by the red circle (b) and the vertical red line (c), respectively.}
\end{figure}

When measuring the momentum distribution for varying densities, see Fig.~\ref{fig:Fig4}(a), we observe that the occupation at low momenta saturates to values close to unity for densities ranging from $\SI{0.25}{\micro\meter^{-2}}$ to $\SI{0.5}{\micro\meter^{-2}}$. This clearly shows Pauli blocking in momentum space \cite{Mukherjee17}.
  
Finally, we realize an interacting homogeneous 2D Fermi gas close to a broad Feshbach resonance \cite{Zuern13} and apply matter wave focusing. We prepare a single-layer attractive 2D Fermi gas at $B=\SI{1020}{\gauss}$ where the ratio of scattering length $a_{\text{3D}}$ to harmonic oscillator length $l_z=\sqrt{\hbar/m\omega_z}$ is $a_{\text{3D}}/l_z=-0.56$.
In contrast to previous experiments which measured the pair momentum distribution by converting pairs into deeply-bound molecules\cite{Ries15}, we measure the momentum distribution of the individual atoms \cite{Regal05}.
This requires a negligible influence of collisions on the time evolution.
We achieve this by releasing the gas from the vertical confinement \cite{Kruger07,Clade09} as well as flashing on a light pulse propagating along the $z$-direction which rapidly ejects atoms in state $|\uparrow\rangle$ \cite{Supp,Weinberg2016,Murmann2015,Mukherjee17}.
This projects the wave function of atoms in state $|\downarrow\rangle$ onto free particle states and allows us to
extract the occupation $f(k)$ of the interacting system using the matter wave focusing technique described above.
For our interaction strength, we expect only small deviations in $f(k)$ compared to the non-interacting system since at $T=0$ the quasiparticle weight $Z$ and the gap $\Delta$ are calculated to be $Z\approx0.9$ \cite{Bloom75} and $\Delta\approx\SI{21}{\percent}$\cite{Fischer14}. We therefore attribute the reduced central occupation and the broadening of the momentum distribution shown in Fig.~\ref{fig:Fig4}(e) to thermal excitations. 

In this letter, we report on the realization of a homogeneous 2D Fermi gas trapped in a box potential.
We locally probe the system by imprinting a step potential using a DMD and thereby measure the EOS of a non-interacting Fermi gas.
Furthermore, we apply matter wave focusing to directly observe Pauli blocking in the momentum distribution of a non-interacting 2D Fermi gas. 
Finally , we demonstrate that the momentum distribution of interacting gases can also be measured and observe a momentum distribution that is qualitatively similar to that of a non-interacting gas for intermediate interactions.

The homogeneous systems presented in this work are particularly useful for studying non-equilibrium dynamics of strongly correlated systems, since they allow interaction quenches without triggering mass redistribution, which is unavoidable in harmonic traps.
The combination of such a homogeneous system with non-local probes is ideally suited to observe critical phenomena and exotic phases such as FFLO superfluidity, which are predicted to exist only in narrow regions of the phase diagram.
Finally, our measurement of the momentum distribution of an interacting Fermi gas can be extended to analyze momentum correlations \cite{Greiner05} and thereby observe Cooper pairs in a fermionic superfluid.

We thank W. Weimer, K. Morgener and K. Riechers for their contributions during earlier stages of the experiment and F. Werner and M.W. Zwierlein for their helpful comments.
This work was supported by the European Union's Seventh Framework Programme (FP7/2007-2013) under grant agreement No. 335431 and by the DFG in the framework of SFB 925, GrK 1355 and the excellence cluster 'The Hamburg Centre for Ultrafast Imaging'.

\bibliography{Box}

\begin{thebibliography}{54}%
\makeatletter
\providecommand \@ifxundefined [1]{%
 \@ifx{#1\undefined}
}%
\providecommand \@ifnum [1]{%
 \ifnum #1\expandafter \@firstoftwo
 \else \expandafter \@secondoftwo
 \fi
}%
\providecommand \@ifx [1]{%
 \ifx #1\expandafter \@firstoftwo
 \else \expandafter \@secondoftwo
 \fi
}%
\providecommand \natexlab [1]{#1}%
\providecommand \enquote  [1]{``#1''}%
\providecommand \bibnamefont  [1]{#1}%
\providecommand \bibfnamefont [1]{#1}%
\providecommand \citenamefont [1]{#1}%
\providecommand \href@noop [0]{\@secondoftwo}%
\providecommand \href [0]{\begingroup \@sanitize@url \@href}%
\providecommand \@href[1]{\@@startlink{#1}\@@href}%
\providecommand \@@href[1]{\endgroup#1\@@endlink}%
\providecommand \@sanitize@url [0]{\catcode `\\12\catcode `\$12\catcode
  `\&12\catcode `\#12\catcode `\^12\catcode `\_12\catcode `\%12\relax}%
\providecommand \@@startlink[1]{}%
\providecommand \@@endlink[0]{}%
\providecommand \url  [0]{\begingroup\@sanitize@url \@url }%
\providecommand \@url [1]{\endgroup\@href {#1}{\urlprefix }}%
\providecommand \urlprefix  [0]{URL }%
\providecommand \Eprint [0]{\href }%
\providecommand \doibase [0]{http://dx.doi.org/}%
\providecommand \selectlanguage [0]{\@gobble}%
\providecommand \bibinfo  [0]{\@secondoftwo}%
\providecommand \bibfield  [0]{\@secondoftwo}%
\providecommand \translation [1]{[#1]}%
\providecommand \BibitemOpen [0]{}%
\providecommand \bibitemStop [0]{}%
\providecommand \bibitemNoStop [0]{.\EOS\space}%
\providecommand \EOS [0]{\spacefactor3000\relax}%
\providecommand \BibitemShut  [1]{\csname bibitem#1\endcsname}%
\let\auto@bib@innerbib\@empty
\bibitem [{\citenamefont {Martiyanov}\ \emph {et~al.}(2010)\citenamefont
  {Martiyanov}, \citenamefont {Makhalov},\ and\ \citenamefont
  {Turlapov}}]{Martiyanov10}%
  \BibitemOpen
  \bibfield  {author} {\bibinfo {author} {\bibfnamefont {K.}~\bibnamefont
  {Martiyanov}}, \bibinfo {author} {\bibfnamefont {V.}~\bibnamefont
  {Makhalov}}, \ and\ \bibinfo {author} {\bibfnamefont {A.}~\bibnamefont
  {Turlapov}},\ }\href {\doibase 10.1103/PhysRevLett.105.030404} {\bibfield
  {journal} {\bibinfo  {journal} {Phys. Rev. Lett.}\ }\textbf {\bibinfo
  {volume} {105}},\ \bibinfo {pages} {030404} (\bibinfo {year}
  {2010})}\BibitemShut {NoStop}%
\bibitem [{\citenamefont {Dyke}\ \emph {et~al.}(2011)\citenamefont {Dyke},
  \citenamefont {Kuhnle}, \citenamefont {Whitlock}, \citenamefont {Hu},
  \citenamefont {Mark}, \citenamefont {Hoinka}, \citenamefont {Lingham},
  \citenamefont {Hannaford},\ and\ \citenamefont {Vale}}]{Dyke11}%
  \BibitemOpen
  \bibfield  {author} {\bibinfo {author} {\bibfnamefont {P.}~\bibnamefont
  {Dyke}}, \bibinfo {author} {\bibfnamefont {E.~D.}\ \bibnamefont {Kuhnle}},
  \bibinfo {author} {\bibfnamefont {S.}~\bibnamefont {Whitlock}}, \bibinfo
  {author} {\bibfnamefont {H.}~\bibnamefont {Hu}}, \bibinfo {author}
  {\bibfnamefont {M.}~\bibnamefont {Mark}}, \bibinfo {author} {\bibfnamefont
  {S.}~\bibnamefont {Hoinka}}, \bibinfo {author} {\bibfnamefont
  {M.}~\bibnamefont {Lingham}}, \bibinfo {author} {\bibfnamefont
  {P.}~\bibnamefont {Hannaford}}, \ and\ \bibinfo {author} {\bibfnamefont
  {C.~J.}\ \bibnamefont {Vale}},\ }\href@noop {} {\bibfield  {journal}
  {\bibinfo  {journal} {Phys. Rev. Lett.}\ }\textbf {\bibinfo {volume} {106}},\
  \bibinfo {pages} {105304} (\bibinfo {year} {2011})}\BibitemShut {NoStop}%
\bibitem [{\citenamefont {Ong}\ \emph {et~al.}(2015)\citenamefont {Ong},
  \citenamefont {Cheng}, \citenamefont {Arakelyan},\ and\ \citenamefont
  {Thomas}}]{Ong15}%
  \BibitemOpen
  \bibfield  {author} {\bibinfo {author} {\bibfnamefont {W.}~\bibnamefont
  {Ong}}, \bibinfo {author} {\bibfnamefont {C.}~\bibnamefont {Cheng}}, \bibinfo
  {author} {\bibfnamefont {I.}~\bibnamefont {Arakelyan}}, \ and\ \bibinfo
  {author} {\bibfnamefont {J.~E.}\ \bibnamefont {Thomas}},\ }\href@noop {}
  {\bibfield  {journal} {\bibinfo  {journal} {Phys. Rev. Lett.}\ }\textbf
  {\bibinfo {volume} {114}},\ \bibinfo {pages} {110403} (\bibinfo {year}
  {2015})}\BibitemShut {NoStop}%
\bibitem [{\citenamefont {Mitra}\ \emph {et~al.}(2016)\citenamefont {Mitra},
  \citenamefont {Brown}, \citenamefont {Schau\ss{}}, \citenamefont {Kondov},\
  and\ \citenamefont {Bakr}}]{Mitra16}%
  \BibitemOpen
  \bibfield  {author} {\bibinfo {author} {\bibfnamefont {D.}~\bibnamefont
  {Mitra}}, \bibinfo {author} {\bibfnamefont {P.~T.}\ \bibnamefont {Brown}},
  \bibinfo {author} {\bibfnamefont {P.}~\bibnamefont {Schau\ss{}}}, \bibinfo
  {author} {\bibfnamefont {S.~S.}\ \bibnamefont {Kondov}}, \ and\ \bibinfo
  {author} {\bibfnamefont {W.~S.}\ \bibnamefont {Bakr}},\ }\href {\doibase
  10.1103/PhysRevLett.117.093601} {\bibfield  {journal} {\bibinfo  {journal}
  {Phys. Rev. Lett.}\ }\textbf {\bibinfo {volume} {117}},\ \bibinfo {pages}
  {093601} (\bibinfo {year} {2016})}\BibitemShut {NoStop}%
\bibitem [{\citenamefont {Fr\"ohlich}\ \emph {et~al.}(2011)\citenamefont
  {Fr\"ohlich}, \citenamefont {Feld}, \citenamefont {Vogt}, \citenamefont
  {Koschorreck}, \citenamefont {Zwerger},\ and\ \citenamefont
  {K\"ohl}}]{Frohlich11}%
  \BibitemOpen
  \bibfield  {author} {\bibinfo {author} {\bibfnamefont {B.}~\bibnamefont
  {Fr\"ohlich}}, \bibinfo {author} {\bibfnamefont {M.}~\bibnamefont {Feld}},
  \bibinfo {author} {\bibfnamefont {E.}~\bibnamefont {Vogt}}, \bibinfo {author}
  {\bibfnamefont {M.}~\bibnamefont {Koschorreck}}, \bibinfo {author}
  {\bibfnamefont {W.}~\bibnamefont {Zwerger}}, \ and\ \bibinfo {author}
  {\bibfnamefont {M.}~\bibnamefont {K\"ohl}},\ }\href@noop {} {\bibfield
  {journal} {\bibinfo  {journal} {Phys. Rev. Lett.}\ }\textbf {\bibinfo
  {volume} {106}},\ \bibinfo {pages} {105301} (\bibinfo {year}
  {2011})}\BibitemShut {NoStop}%
\bibitem [{\citenamefont {Feld}\ \emph {et~al.}(2011)\citenamefont {Feld},
  \citenamefont {Fr\"ohlich}, \citenamefont {Vogt}, \citenamefont
  {Koschorreck},\ and\ \citenamefont {K\"ohl}}]{Feld11}%
  \BibitemOpen
  \bibfield  {author} {\bibinfo {author} {\bibfnamefont {M.}~\bibnamefont
  {Feld}}, \bibinfo {author} {\bibfnamefont {B.}~\bibnamefont {Fr\"ohlich}},
  \bibinfo {author} {\bibfnamefont {E.}~\bibnamefont {Vogt}}, \bibinfo {author}
  {\bibfnamefont {M.}~\bibnamefont {Koschorreck}}, \ and\ \bibinfo {author}
  {\bibfnamefont {M.}~\bibnamefont {K\"ohl}},\ }\href@noop {} {\bibfield
  {journal} {\bibinfo  {journal} {Nature}\ }\textbf {\bibinfo {volume} {480}},\
  \bibinfo {pages} {75} (\bibinfo {year} {2011})}\BibitemShut {NoStop}%
\bibitem [{\citenamefont {Sommer}\ \emph {et~al.}(2012)\citenamefont {Sommer},
  \citenamefont {Cheuk}, \citenamefont {Ku}, \citenamefont {Bakr},\ and\
  \citenamefont {Zwierlein}}]{Sommer12}%
  \BibitemOpen
  \bibfield  {author} {\bibinfo {author} {\bibfnamefont {A.~T.}\ \bibnamefont
  {Sommer}}, \bibinfo {author} {\bibfnamefont {L.~W.}\ \bibnamefont {Cheuk}},
  \bibinfo {author} {\bibfnamefont {M.~J.~H.}\ \bibnamefont {Ku}}, \bibinfo
  {author} {\bibfnamefont {W.~S.}\ \bibnamefont {Bakr}}, \ and\ \bibinfo
  {author} {\bibfnamefont {M.~W.}\ \bibnamefont {Zwierlein}},\ }\href@noop {}
  {\bibfield  {journal} {\bibinfo  {journal} {Phys. Rev. Lett.}\ }\textbf
  {\bibinfo {volume} {108}},\ \bibinfo {pages} {045302} (\bibinfo {year}
  {2012})}\BibitemShut {NoStop}%
\bibitem [{\citenamefont {Cheng}\ \emph {et~al.}(2016)\citenamefont {Cheng},
  \citenamefont {Kangara}, \citenamefont {Arakelyan},\ and\ \citenamefont
  {Thomas}}]{Cheng16}%
  \BibitemOpen
  \bibfield  {author} {\bibinfo {author} {\bibfnamefont {C.}~\bibnamefont
  {Cheng}}, \bibinfo {author} {\bibfnamefont {J.}~\bibnamefont {Kangara}},
  \bibinfo {author} {\bibfnamefont {I.}~\bibnamefont {Arakelyan}}, \ and\
  \bibinfo {author} {\bibfnamefont {J.~E.}\ \bibnamefont {Thomas}},\
  }\href@noop {} {\bibfield  {journal} {\bibinfo  {journal} {Phys. Rev. A}\
  }\textbf {\bibinfo {volume} {94}},\ \bibinfo {pages} {031606(R)} (\bibinfo
  {year} {2016})}\BibitemShut {NoStop}%
\bibitem [{\citenamefont {Fr\"ohlich}\ \emph {et~al.}(2012)\citenamefont
  {Fr\"ohlich}, \citenamefont {Feld}, \citenamefont {Vogt}, \citenamefont
  {Koschorreck}, \citenamefont {K\"ohl}, \citenamefont {Berthod},\ and\
  \citenamefont {Giamarchi}}]{Frohlich12}%
  \BibitemOpen
  \bibfield  {author} {\bibinfo {author} {\bibfnamefont {B.}~\bibnamefont
  {Fr\"ohlich}}, \bibinfo {author} {\bibfnamefont {M.}~\bibnamefont {Feld}},
  \bibinfo {author} {\bibfnamefont {E.}~\bibnamefont {Vogt}}, \bibinfo {author}
  {\bibfnamefont {M.}~\bibnamefont {Koschorreck}}, \bibinfo {author}
  {\bibfnamefont {M.}~\bibnamefont {K\"ohl}}, \bibinfo {author} {\bibfnamefont
  {C.}~\bibnamefont {Berthod}}, \ and\ \bibinfo {author} {\bibfnamefont
  {T.}~\bibnamefont {Giamarchi}},\ }\href@noop {} {\bibfield  {journal}
  {\bibinfo  {journal} {Phys. Rev. Lett.}\ }\textbf {\bibinfo {volume} {109}},\
  \bibinfo {pages} {130403} (\bibinfo {year} {2012})}\BibitemShut {NoStop}%
\bibitem [{\citenamefont {Koschorreck}\ \emph {et~al.}(2012)\citenamefont
  {Koschorreck}, \citenamefont {Pertot}, \citenamefont {Vogt}, \citenamefont
  {Fr\"ohlich}, \citenamefont {Feld},\ and\ \citenamefont
  {K\"ohl}}]{Koschorreck12}%
  \BibitemOpen
  \bibfield  {author} {\bibinfo {author} {\bibfnamefont {M.}~\bibnamefont
  {Koschorreck}}, \bibinfo {author} {\bibfnamefont {D.}~\bibnamefont {Pertot}},
  \bibinfo {author} {\bibfnamefont {E.}~\bibnamefont {Vogt}}, \bibinfo {author}
  {\bibfnamefont {B.}~\bibnamefont {Fr\"ohlich}}, \bibinfo {author}
  {\bibfnamefont {M.}~\bibnamefont {Feld}}, \ and\ \bibinfo {author}
  {\bibfnamefont {M.}~\bibnamefont {K\"ohl}},\ }\href@noop {} {\bibfield
  {journal} {\bibinfo  {journal} {Nature}\ }\textbf {\bibinfo {volume} {485}},\
  \bibinfo {pages} {619} (\bibinfo {year} {2012})}\BibitemShut {NoStop}%
\bibitem [{\citenamefont {Zhang}\ \emph {et~al.}(2012)\citenamefont {Zhang},
  \citenamefont {Ong}, \citenamefont {Arakelyan},\ and\ \citenamefont
  {Thomas}}]{Zhang12}%
  \BibitemOpen
  \bibfield  {author} {\bibinfo {author} {\bibfnamefont {Y.}~\bibnamefont
  {Zhang}}, \bibinfo {author} {\bibfnamefont {W.}~\bibnamefont {Ong}}, \bibinfo
  {author} {\bibfnamefont {I.}~\bibnamefont {Arakelyan}}, \ and\ \bibinfo
  {author} {\bibfnamefont {J.~E.}\ \bibnamefont {Thomas}},\ }\href@noop {}
  {\bibfield  {journal} {\bibinfo  {journal} {Phys. Rev. Lett.}\ }\textbf
  {\bibinfo {volume} {108}},\ \bibinfo {pages} {235302} (\bibinfo {year}
  {2012})}\BibitemShut {NoStop}%
\bibitem [{\citenamefont {Makhalov}\ \emph {et~al.}(2014)\citenamefont
  {Makhalov}, \citenamefont {Martiyanov},\ and\ \citenamefont
  {Turlapov}}]{Makhalov14}%
  \BibitemOpen
  \bibfield  {author} {\bibinfo {author} {\bibfnamefont {V.}~\bibnamefont
  {Makhalov}}, \bibinfo {author} {\bibfnamefont {K.}~\bibnamefont
  {Martiyanov}}, \ and\ \bibinfo {author} {\bibfnamefont {A.}~\bibnamefont
  {Turlapov}},\ }\href {\doibase 10.1103/PhysRevLett.112.045301} {\bibfield
  {journal} {\bibinfo  {journal} {Phys. Rev. Lett.}\ }\textbf {\bibinfo
  {volume} {112}},\ \bibinfo {pages} {045301} (\bibinfo {year}
  {2014})}\BibitemShut {NoStop}%
\bibitem [{\citenamefont {Fenech}\ \emph {et~al.}(2016)\citenamefont {Fenech},
  \citenamefont {Dyke}, \citenamefont {Peppler}, \citenamefont {Lingham},
  \citenamefont {Hoinka}, \citenamefont {Hu},\ and\ \citenamefont
  {Vale}}]{Fenech16}%
  \BibitemOpen
  \bibfield  {author} {\bibinfo {author} {\bibfnamefont {K.}~\bibnamefont
  {Fenech}}, \bibinfo {author} {\bibfnamefont {P.}~\bibnamefont {Dyke}},
  \bibinfo {author} {\bibfnamefont {T.}~\bibnamefont {Peppler}}, \bibinfo
  {author} {\bibfnamefont {M.~G.}\ \bibnamefont {Lingham}}, \bibinfo {author}
  {\bibfnamefont {S.}~\bibnamefont {Hoinka}}, \bibinfo {author} {\bibfnamefont
  {H.}~\bibnamefont {Hu}}, \ and\ \bibinfo {author} {\bibfnamefont {C.~J.}\
  \bibnamefont {Vale}},\ }\href@noop {} {\bibfield  {journal} {\bibinfo
  {journal} {Phys. Rev. Lett.}\ }\textbf {\bibinfo {volume} {116}},\ \bibinfo
  {pages} {045302} (\bibinfo {year} {2016})}\BibitemShut {NoStop}%
\bibitem [{\citenamefont {Boettcher}\ \emph {et~al.}(2016)\citenamefont
  {Boettcher}, \citenamefont {Bayha}, \citenamefont {Kedar}, \citenamefont
  {Murthy}, \citenamefont {Neidig}, \citenamefont {Ries}, \citenamefont {Wenz},
  \citenamefont {Z\"urn}, \citenamefont {Jochim},\ and\ \citenamefont
  {Enss}}]{Boettcher16}%
  \BibitemOpen
  \bibfield  {author} {\bibinfo {author} {\bibfnamefont {I.}~\bibnamefont
  {Boettcher}}, \bibinfo {author} {\bibfnamefont {L.}~\bibnamefont {Bayha}},
  \bibinfo {author} {\bibfnamefont {D.}~\bibnamefont {Kedar}}, \bibinfo
  {author} {\bibfnamefont {P.~A.}\ \bibnamefont {Murthy}}, \bibinfo {author}
  {\bibfnamefont {M.}~\bibnamefont {Neidig}}, \bibinfo {author} {\bibfnamefont
  {M.~G.}\ \bibnamefont {Ries}}, \bibinfo {author} {\bibfnamefont {A.~N.}\
  \bibnamefont {Wenz}}, \bibinfo {author} {\bibfnamefont {G.}~\bibnamefont
  {Z\"urn}}, \bibinfo {author} {\bibfnamefont {S.}~\bibnamefont {Jochim}}, \
  and\ \bibinfo {author} {\bibfnamefont {T.}~\bibnamefont {Enss}},\ }\href
  {\doibase 10.1103/PhysRevLett.116.045303} {\bibfield  {journal} {\bibinfo
  {journal} {Phys. Rev. Lett.}\ }\textbf {\bibinfo {volume} {116}},\ \bibinfo
  {pages} {045303} (\bibinfo {year} {2016})}\BibitemShut {NoStop}%
\bibitem [{\citenamefont {Ries}\ \emph {et~al.}(2015)\citenamefont {Ries},
  \citenamefont {Wenz}, \citenamefont {Z\"urn}, \citenamefont {Bayha},
  \citenamefont {Boettcher}, \citenamefont {Kedar}, \citenamefont {Murthy},
  \citenamefont {Neidig}, \citenamefont {Lompe},\ and\ \citenamefont
  {Jochim}}]{Ries15}%
  \BibitemOpen
  \bibfield  {author} {\bibinfo {author} {\bibfnamefont {M.~G.}\ \bibnamefont
  {Ries}}, \bibinfo {author} {\bibfnamefont {A.~N.}\ \bibnamefont {Wenz}},
  \bibinfo {author} {\bibfnamefont {G.}~\bibnamefont {Z\"urn}}, \bibinfo
  {author} {\bibfnamefont {L.}~\bibnamefont {Bayha}}, \bibinfo {author}
  {\bibfnamefont {I.}~\bibnamefont {Boettcher}}, \bibinfo {author}
  {\bibfnamefont {D.}~\bibnamefont {Kedar}}, \bibinfo {author} {\bibfnamefont
  {P.~A.}\ \bibnamefont {Murthy}}, \bibinfo {author} {\bibfnamefont
  {M.}~\bibnamefont {Neidig}}, \bibinfo {author} {\bibfnamefont
  {T.}~\bibnamefont {Lompe}}, \ and\ \bibinfo {author} {\bibfnamefont
  {S.}~\bibnamefont {Jochim}},\ }\href {\doibase
  10.1103/PhysRevLett.114.230401} {\bibfield  {journal} {\bibinfo  {journal}
  {Phys. Rev. Lett.}\ }\textbf {\bibinfo {volume} {114}},\ \bibinfo {pages}
  {230401} (\bibinfo {year} {2015})}\BibitemShut {NoStop}%
\bibitem [{\citenamefont {Murthy}\ \emph {et~al.}(2015)\citenamefont {Murthy},
  \citenamefont {Boettcher}, \citenamefont {Bayha}, \citenamefont {Holzmann},
  \citenamefont {Kedar}, \citenamefont {Neidig}, \citenamefont {Ries},
  \citenamefont {Wenz}, \citenamefont {Z\"urn},\ and\ \citenamefont
  {Jochim}}]{Murthy15}%
  \BibitemOpen
  \bibfield  {author} {\bibinfo {author} {\bibfnamefont {P.~A.}\ \bibnamefont
  {Murthy}}, \bibinfo {author} {\bibfnamefont {I.}~\bibnamefont {Boettcher}},
  \bibinfo {author} {\bibfnamefont {L.}~\bibnamefont {Bayha}}, \bibinfo
  {author} {\bibfnamefont {M.}~\bibnamefont {Holzmann}}, \bibinfo {author}
  {\bibfnamefont {D.}~\bibnamefont {Kedar}}, \bibinfo {author} {\bibfnamefont
  {M.}~\bibnamefont {Neidig}}, \bibinfo {author} {\bibfnamefont {M.~G.}\
  \bibnamefont {Ries}}, \bibinfo {author} {\bibfnamefont {A.~N.}\ \bibnamefont
  {Wenz}}, \bibinfo {author} {\bibfnamefont {G.}~\bibnamefont {Z\"urn}}, \ and\
  \bibinfo {author} {\bibfnamefont {S.}~\bibnamefont {Jochim}},\ }\href
  {\doibase 10.1103/PhysRevLett.115.010401} {\bibfield  {journal} {\bibinfo
  {journal} {Phys. Rev. Lett.}\ }\textbf {\bibinfo {volume} {115}},\ \bibinfo
  {pages} {010401} (\bibinfo {year} {2015})}\BibitemShut {NoStop}%
\bibitem [{\citenamefont {Fulde}\ and\ \citenamefont
  {Ferrell}(1964)}]{Fulde64}%
  \BibitemOpen
  \bibfield  {author} {\bibinfo {author} {\bibfnamefont {P.}~\bibnamefont
  {Fulde}}\ and\ \bibinfo {author} {\bibfnamefont {R.~A.}\ \bibnamefont
  {Ferrell}},\ }\href {\doibase 10.1103/PhysRev.135.A550} {\bibfield  {journal}
  {\bibinfo  {journal} {Phys. Rev.}\ }\textbf {\bibinfo {volume} {135}},\
  \bibinfo {pages} {A550} (\bibinfo {year} {1964})}\BibitemShut {NoStop}%
\bibitem [{\citenamefont {Larkin}\ and\ \citenamefont
  {Ovchinnikov}(1965)}]{Larkin65}%
  \BibitemOpen
  \bibfield  {author} {\bibinfo {author} {\bibfnamefont {A.}~\bibnamefont
  {Larkin}}\ and\ \bibinfo {author} {\bibfnamefont {Y.}~\bibnamefont
  {Ovchinnikov}},\ }\href@noop {} {\bibfield  {journal} {\bibinfo  {journal}
  {Sov. Phys. JETP}\ }\textbf {\bibinfo {volume} {20}},\ \bibinfo {pages} {762}
  (\bibinfo {year} {1965})}\BibitemShut {NoStop}%
\bibitem [{\citenamefont {Conduit}\ \emph {et~al.}(2008)\citenamefont
  {Conduit}, \citenamefont {Conlon},\ and\ \citenamefont {Simons}}]{Conduit08}%
  \BibitemOpen
  \bibfield  {author} {\bibinfo {author} {\bibfnamefont {G.~J.}\ \bibnamefont
  {Conduit}}, \bibinfo {author} {\bibfnamefont {P.~H.}\ \bibnamefont {Conlon}},
  \ and\ \bibinfo {author} {\bibfnamefont {B.~D.}\ \bibnamefont {Simons}},\
  }\href {\doibase 10.1103/PhysRevA.77.053617} {\bibfield  {journal} {\bibinfo
  {journal} {Phys. Rev. A}\ }\textbf {\bibinfo {volume} {77}},\ \bibinfo
  {pages} {053617} (\bibinfo {year} {2008})}\BibitemShut {NoStop}%
\bibitem [{\citenamefont {Toniolo}\ \emph {et~al.}(2017)\citenamefont
  {Toniolo}, \citenamefont {Mulkerin}, \citenamefont {Liu},\ and\ \citenamefont
  {Hu}}]{Toniolo17}%
  \BibitemOpen
  \bibfield  {author} {\bibinfo {author} {\bibfnamefont {U.}~\bibnamefont
  {Toniolo}}, \bibinfo {author} {\bibfnamefont {B.}~\bibnamefont {Mulkerin}},
  \bibinfo {author} {\bibfnamefont {X.-J.}\ \bibnamefont {Liu}}, \ and\
  \bibinfo {author} {\bibfnamefont {H.}~\bibnamefont {Hu}},\ }\href {\doibase
  10.1103/PhysRevA.95.013603} {\bibfield  {journal} {\bibinfo  {journal} {Phys.
  Rev. A}\ }\textbf {\bibinfo {volume} {95}},\ \bibinfo {pages} {013603}
  (\bibinfo {year} {2017})}\BibitemShut {NoStop}%
\bibitem [{\citenamefont {Gaunt}\ \emph {et~al.}(2013)\citenamefont {Gaunt},
  \citenamefont {Schmidutz}, \citenamefont {Gotlibovych}, \citenamefont
  {Smith},\ and\ \citenamefont {Hadzibabic}}]{Gaunt13}%
  \BibitemOpen
  \bibfield  {author} {\bibinfo {author} {\bibfnamefont {A.~L.}\ \bibnamefont
  {Gaunt}}, \bibinfo {author} {\bibfnamefont {T.~F.}\ \bibnamefont
  {Schmidutz}}, \bibinfo {author} {\bibfnamefont {I.}~\bibnamefont
  {Gotlibovych}}, \bibinfo {author} {\bibfnamefont {R.~P.}\ \bibnamefont
  {Smith}}, \ and\ \bibinfo {author} {\bibfnamefont {Z.}~\bibnamefont
  {Hadzibabic}},\ }\href@noop {} {\bibfield  {journal} {\bibinfo  {journal}
  {Phys. Rev. Lett.}\ }\textbf {\bibinfo {volume} {110}},\ \bibinfo {pages}
  {200406} (\bibinfo {year} {2013})}\BibitemShut {NoStop}%
\bibitem [{\citenamefont {Schmidutz}\ \emph {et~al.}(2014)\citenamefont
  {Schmidutz}, \citenamefont {Gotlibovych}, \citenamefont {Gaunt},
  \citenamefont {Smith}, \citenamefont {Navon},\ and\ \citenamefont
  {Hadzibabic}}]{Schmidutz14}%
  \BibitemOpen
  \bibfield  {author} {\bibinfo {author} {\bibfnamefont {T.~F.}\ \bibnamefont
  {Schmidutz}}, \bibinfo {author} {\bibfnamefont {I.}~\bibnamefont
  {Gotlibovych}}, \bibinfo {author} {\bibfnamefont {A.~L.}\ \bibnamefont
  {Gaunt}}, \bibinfo {author} {\bibfnamefont {R.~P.}\ \bibnamefont {Smith}},
  \bibinfo {author} {\bibfnamefont {N.}~\bibnamefont {Navon}}, \ and\ \bibinfo
  {author} {\bibfnamefont {Z.}~\bibnamefont {Hadzibabic}},\ }\href@noop {}
  {\bibfield  {journal} {\bibinfo  {journal} {Phys. Rev. Lett.}\ }\textbf
  {\bibinfo {volume} {112}},\ \bibinfo {pages} {040403} (\bibinfo {year}
  {2014})}\BibitemShut {NoStop}%
\bibitem [{\citenamefont {Navon}\ \emph {et~al.}(2015)\citenamefont {Navon},
  \citenamefont {Smith},\ and\ \citenamefont {Hadzibabic}}]{Navon15}%
  \BibitemOpen
  \bibfield  {author} {\bibinfo {author} {\bibfnamefont {A.~L.}\ \bibnamefont
  {Navon}, \bibfnamefont {N.~Gaunt}}, \bibinfo {author} {\bibfnamefont {R.~P.}\
  \bibnamefont {Smith}}, \ and\ \bibinfo {author} {\bibfnamefont
  {Z.}~\bibnamefont {Hadzibabic}},\ }\href@noop {} {\bibfield  {journal}
  {\bibinfo  {journal} {Science}\ }\textbf {\bibinfo {volume} {347}},\ \bibinfo
  {pages} {167} (\bibinfo {year} {2015})}\BibitemShut {NoStop}%
\bibitem [{\citenamefont {Corman}\ \emph {et~al.}(2014)\citenamefont {Corman},
  \citenamefont {Chomaz}, \citenamefont {Bienaim\'e}, \citenamefont
  {Desbuquois}, \citenamefont {Weitenberg}, \citenamefont {Nascimb\`ene},
  \citenamefont {Dalibard},\ and\ \citenamefont {Beugnon}}]{Corman14}%
  \BibitemOpen
  \bibfield  {author} {\bibinfo {author} {\bibfnamefont {L.}~\bibnamefont
  {Corman}}, \bibinfo {author} {\bibfnamefont {L.}~\bibnamefont {Chomaz}},
  \bibinfo {author} {\bibfnamefont {T.}~\bibnamefont {Bienaim\'e}}, \bibinfo
  {author} {\bibfnamefont {R.}~\bibnamefont {Desbuquois}}, \bibinfo {author}
  {\bibfnamefont {C.}~\bibnamefont {Weitenberg}}, \bibinfo {author}
  {\bibfnamefont {S.}~\bibnamefont {Nascimb\`ene}}, \bibinfo {author}
  {\bibfnamefont {J.}~\bibnamefont {Dalibard}}, \ and\ \bibinfo {author}
  {\bibfnamefont {J.}~\bibnamefont {Beugnon}},\ }\href@noop {} {\bibfield
  {journal} {\bibinfo  {journal} {Phys. Rev. Lett.}\ }\textbf {\bibinfo
  {volume} {113}},\ \bibinfo {pages} {135302} (\bibinfo {year}
  {2014})}\BibitemShut {NoStop}%
\bibitem [{\citenamefont {Chomaz}\ \emph {et~al.}(2015)\citenamefont {Chomaz},
  \citenamefont {Corman}, \citenamefont {Bienaim\'e}, \citenamefont
  {Desbuquois}, \citenamefont {Weitenberg}, \citenamefont {Nascimb\`ene},
  \citenamefont {Beugnon},\ and\ \citenamefont {Dalibard}}]{Chomaz15}%
  \BibitemOpen
  \bibfield  {author} {\bibinfo {author} {\bibfnamefont {L.}~\bibnamefont
  {Chomaz}}, \bibinfo {author} {\bibfnamefont {L.}~\bibnamefont {Corman}},
  \bibinfo {author} {\bibfnamefont {T.}~\bibnamefont {Bienaim\'e}}, \bibinfo
  {author} {\bibfnamefont {R.}~\bibnamefont {Desbuquois}}, \bibinfo {author}
  {\bibfnamefont {C.}~\bibnamefont {Weitenberg}}, \bibinfo {author}
  {\bibfnamefont {S.}~\bibnamefont {Nascimb\`ene}}, \bibinfo {author}
  {\bibfnamefont {J.}~\bibnamefont {Beugnon}}, \ and\ \bibinfo {author}
  {\bibfnamefont {J.}~\bibnamefont {Dalibard}},\ }\href@noop {} {\bibfield
  {journal} {\bibinfo  {journal} {Nat. Commun.}\ }\textbf {\bibinfo {volume}
  {6}},\ \bibinfo {pages} {6162} (\bibinfo {year} {2015})}\BibitemShut
  {NoStop}%
\bibitem [{\citenamefont {Mukherjee}\ \emph {et~al.}(2017)\citenamefont
  {Mukherjee}, \citenamefont {Yan}, \citenamefont {Patel}, \citenamefont
  {Hadzibabic}, \citenamefont {Yefsah}, \citenamefont {Struck},\ and\
  \citenamefont {Zwierlein}}]{Mukherjee17}%
  \BibitemOpen
  \bibfield  {author} {\bibinfo {author} {\bibfnamefont {B.}~\bibnamefont
  {Mukherjee}}, \bibinfo {author} {\bibfnamefont {Z.}~\bibnamefont {Yan}},
  \bibinfo {author} {\bibfnamefont {P.~B.}\ \bibnamefont {Patel}}, \bibinfo
  {author} {\bibfnamefont {Z.}~\bibnamefont {Hadzibabic}}, \bibinfo {author}
  {\bibfnamefont {T.}~\bibnamefont {Yefsah}}, \bibinfo {author} {\bibfnamefont
  {J.}~\bibnamefont {Struck}}, \ and\ \bibinfo {author} {\bibfnamefont {M.~W.}\
  \bibnamefont {Zwierlein}},\ }\href {\doibase 10.1103/PhysRevLett.118.123401}
  {\bibfield  {journal} {\bibinfo  {journal} {Phys. Rev. Lett.}\ }\textbf
  {\bibinfo {volume} {118}},\ \bibinfo {pages} {123401} (\bibinfo {year}
  {2017})}\BibitemShut {NoStop}%
\bibitem [{\citenamefont {Weimer}\ \emph {et~al.}(2015)\citenamefont {Weimer},
  \citenamefont {Morgener}, \citenamefont {Singh}, \citenamefont {Siegl},
  \citenamefont {Hueck}, \citenamefont {Luick}, \citenamefont {Mathey},\ and\
  \citenamefont {Moritz}}]{Weimer15}%
  \BibitemOpen
  \bibfield  {author} {\bibinfo {author} {\bibfnamefont {W.}~\bibnamefont
  {Weimer}}, \bibinfo {author} {\bibfnamefont {K.}~\bibnamefont {Morgener}},
  \bibinfo {author} {\bibfnamefont {V.~P.}\ \bibnamefont {Singh}}, \bibinfo
  {author} {\bibfnamefont {J.}~\bibnamefont {Siegl}}, \bibinfo {author}
  {\bibfnamefont {K.}~\bibnamefont {Hueck}}, \bibinfo {author} {\bibfnamefont
  {N.}~\bibnamefont {Luick}}, \bibinfo {author} {\bibfnamefont
  {L.}~\bibnamefont {Mathey}}, \ and\ \bibinfo {author} {\bibfnamefont
  {H.}~\bibnamefont {Moritz}},\ }\href {\doibase
  10.1103/PhysRevLett.114.095301} {\bibfield  {journal} {\bibinfo  {journal}
  {Phys. Rev. Lett.}\ }\textbf {\bibinfo {volume} {114}},\ \bibinfo {pages}
  {095301} (\bibinfo {year} {2015})}\BibitemShut {NoStop}%
\bibitem [{Sup()}]{Supp}%
  \BibitemOpen
  \href@noop {} {}\bibinfo {note} {See Supplemental Material}\BibitemShut
  {NoStop}%
\bibitem [{\citenamefont {McLeod}(1954)}]{McLeod54}%
  \BibitemOpen
  \bibfield  {author} {\bibinfo {author} {\bibfnamefont {J.~H.}\ \bibnamefont
  {McLeod}},\ }\href {\doibase 10.1364/JOSA.44.000592} {\bibfield  {journal}
  {\bibinfo  {journal} {J. Opt. Soc. Am.}\ }\textbf {\bibinfo {volume} {44}},\
  \bibinfo {pages} {592} (\bibinfo {year} {1954})}\BibitemShut {NoStop}%
\bibitem [{\citenamefont {Manek}\ \emph {et~al.}(1998)\citenamefont {Manek},
  \citenamefont {Ovchinnikov},\ and\ \citenamefont {Grimm}}]{Manek98}%
  \BibitemOpen
  \bibfield  {author} {\bibinfo {author} {\bibfnamefont {I.}~\bibnamefont
  {Manek}}, \bibinfo {author} {\bibfnamefont {Y.}~\bibnamefont {Ovchinnikov}},
  \ and\ \bibinfo {author} {\bibfnamefont {R.}~\bibnamefont {Grimm}},\ }\href
  {\doibase http://dx.doi.org/10.1016/S0030-4018(97)00645-7} {\bibfield
  {journal} {\bibinfo  {journal} {Opt. Commun.}\ }\textbf {\bibinfo {volume}
  {147}},\ \bibinfo {pages} {67 } (\bibinfo {year} {1998})}\BibitemShut
  {NoStop}%
\bibitem [{\citenamefont {Petrov}\ \emph {et~al.}(2000)\citenamefont {Petrov},
  \citenamefont {Holzmann},\ and\ \citenamefont {Shlyapnikov}}]{Petrov00}%
  \BibitemOpen
  \bibfield  {author} {\bibinfo {author} {\bibfnamefont {D.~S.}\ \bibnamefont
  {Petrov}}, \bibinfo {author} {\bibfnamefont {M.}~\bibnamefont {Holzmann}}, \
  and\ \bibinfo {author} {\bibfnamefont {G.~V.}\ \bibnamefont {Shlyapnikov}},\
  }\href {\doibase 10.1103/PhysRevLett.84.2551} {\bibfield  {journal} {\bibinfo
   {journal} {Phys. Rev. Lett.}\ }\textbf {\bibinfo {volume} {84}},\ \bibinfo
  {pages} {2551} (\bibinfo {year} {2000})}\BibitemShut {NoStop}%
\bibitem [{\citenamefont {Dyke}\ \emph {et~al.}(2016)\citenamefont {Dyke},
  \citenamefont {Fenech}, \citenamefont {Peppler}, \citenamefont {Lingham},
  \citenamefont {Hoinka}, \citenamefont {Zhang}, \citenamefont {Peng},
  \citenamefont {Mulkerin}, \citenamefont {Hu}, \citenamefont {Liu},\ and\
  \citenamefont {Vale}}]{Dyke16}%
  \BibitemOpen
  \bibfield  {author} {\bibinfo {author} {\bibfnamefont {P.}~\bibnamefont
  {Dyke}}, \bibinfo {author} {\bibfnamefont {K.}~\bibnamefont {Fenech}},
  \bibinfo {author} {\bibfnamefont {T.}~\bibnamefont {Peppler}}, \bibinfo
  {author} {\bibfnamefont {M.~G.}\ \bibnamefont {Lingham}}, \bibinfo {author}
  {\bibfnamefont {S.}~\bibnamefont {Hoinka}}, \bibinfo {author} {\bibfnamefont
  {W.}~\bibnamefont {Zhang}}, \bibinfo {author} {\bibfnamefont {S.-G.}\
  \bibnamefont {Peng}}, \bibinfo {author} {\bibfnamefont {B.}~\bibnamefont
  {Mulkerin}}, \bibinfo {author} {\bibfnamefont {H.}~\bibnamefont {Hu}},
  \bibinfo {author} {\bibfnamefont {X.-J.}\ \bibnamefont {Liu}}, \ and\
  \bibinfo {author} {\bibfnamefont {C.~J.}\ \bibnamefont {Vale}},\ }\href
  {\doibase 10.1103/PhysRevA.93.011603} {\bibfield  {journal} {\bibinfo
  {journal} {Phys. Rev. A}\ }\textbf {\bibinfo {volume} {93}},\ \bibinfo
  {pages} {011603} (\bibinfo {year} {2016})}\BibitemShut {NoStop}%
\bibitem [{\citenamefont {Hadzibabic}\ \emph {et~al.}(2006)\citenamefont
  {Hadzibabic}, \citenamefont {Kr\"uger}, \citenamefont {Cheneau},
  \citenamefont {Battelier},\ and\ \citenamefont {Dalibard}}]{Hadzibabic06}%
  \BibitemOpen
  \bibfield  {author} {\bibinfo {author} {\bibfnamefont {Z.}~\bibnamefont
  {Hadzibabic}}, \bibinfo {author} {\bibfnamefont {P.}~\bibnamefont
  {Kr\"uger}}, \bibinfo {author} {\bibfnamefont {M.}~\bibnamefont {Cheneau}},
  \bibinfo {author} {\bibfnamefont {B.}~\bibnamefont {Battelier}}, \ and\
  \bibinfo {author} {\bibfnamefont {J.}~\bibnamefont {Dalibard}},\ }\href@noop
  {} {\bibfield  {journal} {\bibinfo  {journal} {Nature}\ }\textbf {\bibinfo
  {volume} {441}},\ \bibinfo {pages} {1118} (\bibinfo {year}
  {2006})}\BibitemShut {NoStop}%
\bibitem [{\citenamefont {Hueck}\ \emph
  {et~al.}(2017{\natexlab{a}})\citenamefont {Hueck}, \citenamefont {Mazurenko},
  \citenamefont {Luick}, \citenamefont {Lompe},\ and\ \citenamefont
  {Moritz}}]{Hueck17}%
  \BibitemOpen
  \bibfield  {author} {\bibinfo {author} {\bibfnamefont {K.}~\bibnamefont
  {Hueck}}, \bibinfo {author} {\bibfnamefont {A.}~\bibnamefont {Mazurenko}},
  \bibinfo {author} {\bibfnamefont {N.}~\bibnamefont {Luick}}, \bibinfo
  {author} {\bibfnamefont {T.}~\bibnamefont {Lompe}}, \ and\ \bibinfo {author}
  {\bibfnamefont {H.}~\bibnamefont {Moritz}},\ }\href {\doibase
  10.1063/1.4973969} {\bibfield  {journal} {\bibinfo  {journal} {Rev. Sci.
  Instrum.}\ }\textbf {\bibinfo {volume} {88}},\ \bibinfo {pages} {016103}
  (\bibinfo {year} {2017}{\natexlab{a}})}\BibitemShut {NoStop}%
\bibitem [{\citenamefont {Reinaudi}\ \emph {et~al.}(2007)\citenamefont
  {Reinaudi}, \citenamefont {Lahaye}, \citenamefont {Wang},\ and\ \citenamefont
  {Gu\'ery-Odelin}}]{Reinaudi07}%
  \BibitemOpen
  \bibfield  {author} {\bibinfo {author} {\bibfnamefont {G.}~\bibnamefont
  {Reinaudi}}, \bibinfo {author} {\bibfnamefont {T.}~\bibnamefont {Lahaye}},
  \bibinfo {author} {\bibfnamefont {Z.}~\bibnamefont {Wang}}, \ and\ \bibinfo
  {author} {\bibfnamefont {D.}~\bibnamefont {Gu\'ery-Odelin}},\ }\href@noop {}
  {\bibfield  {journal} {\bibinfo  {journal} {Opt. Lett.}\ }\textbf {\bibinfo
  {volume} {32}},\ \bibinfo {pages} {3143} (\bibinfo {year}
  {2007})}\BibitemShut {NoStop}%
\bibitem [{\citenamefont {Hueck}\ \emph
  {et~al.}(2017{\natexlab{b}})\citenamefont {Hueck}, \citenamefont {Luick},
  \citenamefont {Sobirey}, \citenamefont {Lompe}, \citenamefont {Moritz},
  \citenamefont {Clark},\ and\ \citenamefont {Chin}}]{Hueck171}%
  \BibitemOpen
  \bibfield  {author} {\bibinfo {author} {\bibfnamefont {K.}~\bibnamefont
  {Hueck}}, \bibinfo {author} {\bibfnamefont {N.}~\bibnamefont {Luick}},
  \bibinfo {author} {\bibfnamefont {L.}~\bibnamefont {Sobirey}}, \bibinfo
  {author} {\bibfnamefont {T.}~\bibnamefont {Lompe}}, \bibinfo {author}
  {\bibfnamefont {H.}~\bibnamefont {Moritz}}, \bibinfo {author} {\bibfnamefont
  {L.}~\bibnamefont {Clark}}, \ and\ \bibinfo {author} {\bibfnamefont
  {C.}~\bibnamefont {Chin}},\ }\href@noop {} {\bibfield  {journal} {\bibinfo
  {journal} {Opt. Exp.}\ }\textbf {\bibinfo {volume} {25}} (\bibinfo {year}
  {2017}{\natexlab{b}})}\BibitemShut {NoStop}%
\bibitem [{\citenamefont {Bauer}\ \emph {et~al.}(2014)\citenamefont {Bauer},
  \citenamefont {Parish},\ and\ \citenamefont {Enss}}]{Bauer14}%
  \BibitemOpen
  \bibfield  {author} {\bibinfo {author} {\bibfnamefont {M.}~\bibnamefont
  {Bauer}}, \bibinfo {author} {\bibfnamefont {M.~M.}\ \bibnamefont {Parish}}, \
  and\ \bibinfo {author} {\bibfnamefont {T.}~\bibnamefont {Enss}},\ }\href
  {\doibase 10.1103/PhysRevLett.112.135302} {\bibfield  {journal} {\bibinfo
  {journal} {Phys. Rev. Lett.}\ }\textbf {\bibinfo {volume} {112}},\ \bibinfo
  {pages} {135302} (\bibinfo {year} {2014})}\BibitemShut {NoStop}%
\bibitem [{Note1()}]{Note1}%
  \BibitemOpen
  \bibinfo {note} {This equation is found by solving the 2D EOS for $\mu _0$,
  taking the $T=0$ limit, yielding $T_F=n_{\protect \text {2D},\delimiter
  "3222378 }\protect \tmspace +\thinmuskip {.1667em}2\pi \hbar ^2/(m k_B) $ and
  reinserting the EOS for $n_{\protect \text {2D},\delimiter "3222378
  }$}\BibitemShut {NoStop}%
\bibitem [{\citenamefont {Hung}\ \emph {et~al.}(2011)\citenamefont {Hung},
  \citenamefont {Zhang}, \citenamefont {Gemelke},\ and\ \citenamefont
  {Chin}}]{Hung11}%
  \BibitemOpen
  \bibfield  {author} {\bibinfo {author} {\bibfnamefont {C.-L.}\ \bibnamefont
  {Hung}}, \bibinfo {author} {\bibfnamefont {X.}~\bibnamefont {Zhang}},
  \bibinfo {author} {\bibfnamefont {N.}~\bibnamefont {Gemelke}}, \ and\
  \bibinfo {author} {\bibfnamefont {C.}~\bibnamefont {Chin}},\ }\href@noop {}
  {\bibfield  {journal} {\bibinfo  {journal} {Nature}\ }\textbf {\bibinfo
  {volume} {470}},\ \bibinfo {pages} {236} (\bibinfo {year}
  {2011})}\BibitemShut {NoStop}%
\bibitem [{\citenamefont {Shvarchuck}\ \emph {et~al.}(2002)\citenamefont
  {Shvarchuck}, \citenamefont {Buggle}, \citenamefont {Petrov}, \citenamefont
  {Dieckmann}, \citenamefont {Zielonkovski}, \citenamefont {Kemmann},
  \citenamefont {Tiecke}, \citenamefont {von Klitzing}, \citenamefont
  {Shlyapnikov},\ and\ \citenamefont {Walraven}}]{Shvarchuck02}%
  \BibitemOpen
  \bibfield  {author} {\bibinfo {author} {\bibfnamefont {I.}~\bibnamefont
  {Shvarchuck}}, \bibinfo {author} {\bibfnamefont {C.}~\bibnamefont {Buggle}},
  \bibinfo {author} {\bibfnamefont {D.~S.}\ \bibnamefont {Petrov}}, \bibinfo
  {author} {\bibfnamefont {K.}~\bibnamefont {Dieckmann}}, \bibinfo {author}
  {\bibfnamefont {M.}~\bibnamefont {Zielonkovski}}, \bibinfo {author}
  {\bibfnamefont {M.}~\bibnamefont {Kemmann}}, \bibinfo {author} {\bibfnamefont
  {T.~G.}\ \bibnamefont {Tiecke}}, \bibinfo {author} {\bibfnamefont
  {W.}~\bibnamefont {von Klitzing}}, \bibinfo {author} {\bibfnamefont {G.~V.}\
  \bibnamefont {Shlyapnikov}}, \ and\ \bibinfo {author} {\bibfnamefont
  {J.~T.~M.}\ \bibnamefont {Walraven}},\ }\href@noop {} {\bibfield  {journal}
  {\bibinfo  {journal} {Phys. Rev. Lett.}\ }\textbf {\bibinfo {volume} {89}},\
  \bibinfo {pages} {270404} (\bibinfo {year} {2002})}\BibitemShut {NoStop}%
\bibitem [{\citenamefont {Tung}\ \emph {et~al.}(2010)\citenamefont {Tung},
  \citenamefont {Lamporesi}, \citenamefont {Lobser}, \citenamefont {Xia},\ and\
  \citenamefont {Cornell}}]{Tung10}%
  \BibitemOpen
  \bibfield  {author} {\bibinfo {author} {\bibfnamefont {S.}~\bibnamefont
  {Tung}}, \bibinfo {author} {\bibfnamefont {G.}~\bibnamefont {Lamporesi}},
  \bibinfo {author} {\bibfnamefont {D.}~\bibnamefont {Lobser}}, \bibinfo
  {author} {\bibfnamefont {L.}~\bibnamefont {Xia}}, \ and\ \bibinfo {author}
  {\bibfnamefont {E.~A.}\ \bibnamefont {Cornell}},\ }\href@noop {} {\bibfield
  {journal} {\bibinfo  {journal} {Phys. Rev. Lett.}\ }\textbf {\bibinfo
  {volume} {105}},\ \bibinfo {pages} {230408} (\bibinfo {year}
  {2010})}\BibitemShut {NoStop}%
\bibitem [{\citenamefont {Murthy}\ \emph {et~al.}(2014)\citenamefont {Murthy},
  \citenamefont {Kedar}, \citenamefont {Lompe}, \citenamefont {Neidig},
  \citenamefont {Ries}, \citenamefont {Wenz}, \citenamefont {Z\"urn},\ and\
  \citenamefont {Jochim}}]{Murthy14}%
  \BibitemOpen
  \bibfield  {author} {\bibinfo {author} {\bibfnamefont {P.~A.}\ \bibnamefont
  {Murthy}}, \bibinfo {author} {\bibfnamefont {D.}~\bibnamefont {Kedar}},
  \bibinfo {author} {\bibfnamefont {T.}~\bibnamefont {Lompe}}, \bibinfo
  {author} {\bibfnamefont {M.}~\bibnamefont {Neidig}}, \bibinfo {author}
  {\bibfnamefont {M.~G.}\ \bibnamefont {Ries}}, \bibinfo {author}
  {\bibfnamefont {A.~N.}\ \bibnamefont {Wenz}}, \bibinfo {author}
  {\bibfnamefont {G.}~\bibnamefont {Z\"urn}}, \ and\ \bibinfo {author}
  {\bibfnamefont {S.}~\bibnamefont {Jochim}},\ }\href {\doibase
  10.1103/PhysRevA.90.043611} {\bibfield  {journal} {\bibinfo  {journal} {Phys.
  Rev. A}\ }\textbf {\bibinfo {volume} {90}},\ \bibinfo {pages} {043611}
  (\bibinfo {year} {2014})}\BibitemShut {NoStop}%
\bibitem [{Note2()}]{Note2}%
  \BibitemOpen
  \bibinfo {note} {In our case this imaging has a magnification of 1. Other
  magnifications are accessible by switching to a different radial trap
  frequency after the $\tau /4$ point.}\BibitemShut {Stop}%
\bibitem [{Note3()}]{Note3}%
  \BibitemOpen
  \bibinfo {note} {We chose a low evaporation depth, despite the fact that we
  achieve our lowest $T/T_F$ at higher Fermi energies, since we want $k_F$ to
  be small enough that the full momentum distribution is captured by the field
  of view of the imaging system.}\BibitemShut {Stop}%
\bibitem [{\citenamefont {Z\"urn}\ \emph {et~al.}(2013)\citenamefont {Z\"urn},
  \citenamefont {Lompe}, \citenamefont {Wenz}, \citenamefont {Jochim},
  \citenamefont {Julienne},\ and\ \citenamefont {Hutson}}]{Zuern13}%
  \BibitemOpen
  \bibfield  {author} {\bibinfo {author} {\bibfnamefont {G.}~\bibnamefont
  {Z\"urn}}, \bibinfo {author} {\bibfnamefont {T.}~\bibnamefont {Lompe}},
  \bibinfo {author} {\bibfnamefont {A.~N.}\ \bibnamefont {Wenz}}, \bibinfo
  {author} {\bibfnamefont {S.}~\bibnamefont {Jochim}}, \bibinfo {author}
  {\bibfnamefont {P.~S.}\ \bibnamefont {Julienne}}, \ and\ \bibinfo {author}
  {\bibfnamefont {J.~M.}\ \bibnamefont {Hutson}},\ }\href {\doibase
  10.1103/PhysRevLett.110.135301} {\bibfield  {journal} {\bibinfo  {journal}
  {Phys. Rev. Lett.}\ }\textbf {\bibinfo {volume} {110}},\ \bibinfo {pages}
  {135301} (\bibinfo {year} {2013})}\BibitemShut {NoStop}%
\bibitem [{\citenamefont {Regal}\ \emph {et~al.}(2005)\citenamefont {Regal},
  \citenamefont {Greiner}, \citenamefont {Giorgini}, \citenamefont {Holland},\
  and\ \citenamefont {Jin}}]{Regal05}%
  \BibitemOpen
  \bibfield  {author} {\bibinfo {author} {\bibfnamefont {C.~A.}\ \bibnamefont
  {Regal}}, \bibinfo {author} {\bibfnamefont {M.}~\bibnamefont {Greiner}},
  \bibinfo {author} {\bibfnamefont {S.}~\bibnamefont {Giorgini}}, \bibinfo
  {author} {\bibfnamefont {M.}~\bibnamefont {Holland}}, \ and\ \bibinfo
  {author} {\bibfnamefont {D.~S.}\ \bibnamefont {Jin}},\ }\href {\doibase
  10.1103/PhysRevLett.95.250404} {\bibfield  {journal} {\bibinfo  {journal}
  {Phys. Rev. Lett.}\ }\textbf {\bibinfo {volume} {95}},\ \bibinfo {pages}
  {250404} (\bibinfo {year} {2005})}\BibitemShut {NoStop}%
\bibitem [{\citenamefont {Kr\"uger}\ \emph {et~al.}(2007)\citenamefont
  {Kr\"uger}, \citenamefont {Hadzibabic},\ and\ \citenamefont
  {Dalibard}}]{Kruger07}%
  \BibitemOpen
  \bibfield  {author} {\bibinfo {author} {\bibfnamefont {P.}~\bibnamefont
  {Kr\"uger}}, \bibinfo {author} {\bibfnamefont {Z.}~\bibnamefont
  {Hadzibabic}}, \ and\ \bibinfo {author} {\bibfnamefont {J.}~\bibnamefont
  {Dalibard}},\ }\href@noop {} {\bibfield  {journal} {\bibinfo  {journal}
  {Phys. Rev. Lett.}\ }\textbf {\bibinfo {volume} {99}},\ \bibinfo {pages}
  {040402} (\bibinfo {year} {2007})}\BibitemShut {NoStop}%
\bibitem [{\citenamefont {Clad\'e}\ \emph {et~al.}(2009)\citenamefont
  {Clad\'e}, \citenamefont {Ryu}, \citenamefont {Ramanathan}, \citenamefont
  {Helmerson},\ and\ \citenamefont {Phillips}}]{Clade09}%
  \BibitemOpen
  \bibfield  {author} {\bibinfo {author} {\bibfnamefont {P.}~\bibnamefont
  {Clad\'e}}, \bibinfo {author} {\bibfnamefont {C.}~\bibnamefont {Ryu}},
  \bibinfo {author} {\bibfnamefont {A.}~\bibnamefont {Ramanathan}}, \bibinfo
  {author} {\bibfnamefont {K.}~\bibnamefont {Helmerson}}, \ and\ \bibinfo
  {author} {\bibfnamefont {W.~D.}\ \bibnamefont {Phillips}},\ }\href@noop {}
  {\bibfield  {journal} {\bibinfo  {journal} {Phys. Rev. Lett.}\ }\textbf
  {\bibinfo {volume} {102}},\ \bibinfo {pages} {170401} (\bibinfo {year}
  {2009})}\BibitemShut {NoStop}%
\bibitem [{\citenamefont {Weinberg}\ \emph {et~al.}(2016)\citenamefont
  {Weinberg}, \citenamefont {J{\"u}rgensen}, \citenamefont
  {{\"O}lschl{\"a}ger}, \citenamefont {L{\"u}hmann}, \citenamefont
  {Sengstock},\ and\ \citenamefont {Simonet}}]{Weinberg2016}%
  \BibitemOpen
  \bibfield  {author} {\bibinfo {author} {\bibfnamefont {M.}~\bibnamefont
  {Weinberg}}, \bibinfo {author} {\bibfnamefont {O.}~\bibnamefont
  {J{\"u}rgensen}}, \bibinfo {author} {\bibfnamefont {C.}~\bibnamefont
  {{\"O}lschl{\"a}ger}}, \bibinfo {author} {\bibfnamefont {D.-S.}\ \bibnamefont
  {L{\"u}hmann}}, \bibinfo {author} {\bibfnamefont {K.}~\bibnamefont
  {Sengstock}}, \ and\ \bibinfo {author} {\bibfnamefont {J.}~\bibnamefont
  {Simonet}},\ }\href@noop {} {\bibfield  {journal} {\bibinfo  {journal}
  {Physical Review A}\ }\textbf {\bibinfo {volume} {93}},\ \bibinfo {pages}
  {033625} (\bibinfo {year} {2016})}\BibitemShut {NoStop}%
\bibitem [{\citenamefont {Murmann}\ \emph {et~al.}(2015)\citenamefont
  {Murmann}, \citenamefont {Deuretzbacher}, \citenamefont {Z{\"u}rn},
  \citenamefont {Bjerlin}, \citenamefont {Reimann}, \citenamefont {Santos},
  \citenamefont {Lompe},\ and\ \citenamefont {Jochim}}]{Murmann2015}%
  \BibitemOpen
  \bibfield  {author} {\bibinfo {author} {\bibfnamefont {S.}~\bibnamefont
  {Murmann}}, \bibinfo {author} {\bibfnamefont {F.}~\bibnamefont
  {Deuretzbacher}}, \bibinfo {author} {\bibfnamefont {G.}~\bibnamefont
  {Z{\"u}rn}}, \bibinfo {author} {\bibfnamefont {J.}~\bibnamefont {Bjerlin}},
  \bibinfo {author} {\bibfnamefont {S.~M.}\ \bibnamefont {Reimann}}, \bibinfo
  {author} {\bibfnamefont {L.}~\bibnamefont {Santos}}, \bibinfo {author}
  {\bibfnamefont {T.}~\bibnamefont {Lompe}}, \ and\ \bibinfo {author}
  {\bibfnamefont {S.}~\bibnamefont {Jochim}},\ }\href@noop {} {\bibfield
  {journal} {\bibinfo  {journal} {Phys. Rev. Lett.}\ }\textbf {\bibinfo
  {volume} {115}},\ \bibinfo {pages} {215301} (\bibinfo {year}
  {2015})}\BibitemShut {NoStop}%
\bibitem [{\citenamefont {Bloom}(1975)}]{Bloom75}%
  \BibitemOpen
  \bibfield  {author} {\bibinfo {author} {\bibfnamefont {P.}~\bibnamefont
  {Bloom}},\ }\href {\doibase 10.1103/PhysRevB.12.125} {\bibfield  {journal}
  {\bibinfo  {journal} {Phys. Rev. B}\ }\textbf {\bibinfo {volume} {12}},\
  \bibinfo {pages} {125} (\bibinfo {year} {1975})}\BibitemShut {NoStop}%
\bibitem [{\citenamefont {Fischer}\ and\ \citenamefont
  {Parish}(2014)}]{Fischer14}%
  \BibitemOpen
  \bibfield  {author} {\bibinfo {author} {\bibfnamefont {A.~M.}\ \bibnamefont
  {Fischer}}\ and\ \bibinfo {author} {\bibfnamefont {M.~M.}\ \bibnamefont
  {Parish}},\ }\href {\doibase 10.1103/PhysRevB.90.214503} {\bibfield
  {journal} {\bibinfo  {journal} {Phys. Rev. B}\ }\textbf {\bibinfo {volume}
  {90}},\ \bibinfo {pages} {214503} (\bibinfo {year} {2014})}\BibitemShut
  {NoStop}%
\bibitem [{\citenamefont {Greiner}\ \emph {et~al.}(2005)\citenamefont
  {Greiner}, \citenamefont {Regal}, \citenamefont {Stewart},\ and\
  \citenamefont {Jin}}]{Greiner05}%
  \BibitemOpen
  \bibfield  {author} {\bibinfo {author} {\bibfnamefont {M.}~\bibnamefont
  {Greiner}}, \bibinfo {author} {\bibfnamefont {C.~A.}\ \bibnamefont {Regal}},
  \bibinfo {author} {\bibfnamefont {J.~T.}\ \bibnamefont {Stewart}}, \ and\
  \bibinfo {author} {\bibfnamefont {D.~S.}\ \bibnamefont {Jin}},\ }\href@noop
  {} {\bibfield  {journal} {\bibinfo  {journal} {Phys. Rev. Lett.}\ }\textbf
  {\bibinfo {volume} {94}},\ \bibinfo {pages} {110401} (\bibinfo {year}
  {2005})}\BibitemShut {NoStop}%
\bibitem [{\citenamefont {Pappa}\ \emph {et~al.}(2011)\citenamefont {Pappa},
  \citenamefont {Condylis}, \citenamefont {Konstantinidis}, \citenamefont
  {Bolpasi}, \citenamefont {Lazoudis}, \citenamefont {Morizot}, \citenamefont
  {Sahagun}, \citenamefont {Baker},\ and\ \citenamefont {von
  Klitzing}}]{Pappa11}%
  \BibitemOpen
  \bibfield  {author} {\bibinfo {author} {\bibfnamefont {M.}~\bibnamefont
  {Pappa}}, \bibinfo {author} {\bibfnamefont {P.~C.}\ \bibnamefont {Condylis}},
  \bibinfo {author} {\bibfnamefont {G.~O.}\ \bibnamefont {Konstantinidis}},
  \bibinfo {author} {\bibfnamefont {V.}~\bibnamefont {Bolpasi}}, \bibinfo
  {author} {\bibfnamefont {A.}~\bibnamefont {Lazoudis}}, \bibinfo {author}
  {\bibfnamefont {O.}~\bibnamefont {Morizot}}, \bibinfo {author} {\bibfnamefont
  {D.}~\bibnamefont {Sahagun}}, \bibinfo {author} {\bibfnamefont
  {M.}~\bibnamefont {Baker}}, \ and\ \bibinfo {author} {\bibfnamefont
  {W.}~\bibnamefont {von Klitzing}},\ }\href
  {http://stacks.iop.org/1367-2630/13/i=11/a=115012} {\bibfield  {journal}
  {\bibinfo  {journal} {New J. Phys.}\ }\textbf {\bibinfo {volume} {13}},\
  \bibinfo {pages} {115012} (\bibinfo {year} {2011})}\BibitemShut {NoStop}%
\end{thebibliography}%

\clearpage
\pagebreak
\begin{center}
\textbf{\large Supplemental Material: Two-Dimensional Homogeneous Fermi Gases}
\end{center}
\setcounter{figure}{0}
\makeatletter
\renewcommand{\thefigure}{S\arabic{figure}}

\section{Trap Geometries}
The elliptic trap is formed by a highly elliptical laser beam with a wavelength of $\lambda = \SI{1064}{\nano\meter}$ and vertical and horizontal beam waists of approximately $w_z \approx \SI{10}{\micro\meter}$ and $w_y\approx\SI{400}{\micro\meter}$. 
At a power of \SI{1}{\watt} this results in trapping frequencies of $(\omega_x,\omega_y,\omega_z) \approx 2\pi \cdot (75,100,4000) \si{\hertz}$.

The hybrid box potential is formed by a superposition of several independently tunable optical and magnetic potentials, which allow for numerous different trapping geometries. 
In all configurations, the $z$-confinement with $\hbar\omega_z = h \cdot \SI{12.4\pm0.1}{\kilo\hertz}$ is realized by the lattice in $z$-direction, which is formed by two $\lambda = \SI{532}{\nano\meter}$ beams intersecting with an opening-angle of \SI{10.4}{\degree} leading to a lattice spacing of \SI{2.9}{\micro\meter}.
This confinement in $z$-direction comes with a radial anti-confinement of $\omega_{x,opt}\approx 2\pi \cdot i\SI{9.5+-1.2}{\hertz}$ and $\omega_{y,opt}\approx 2\pi \cdot i\SI{12.6+-0.8}{\hertz}$.

The main contribution to the radial confinement is typically given by the repulsive ring potential, which is described in detail in the following section.
An additional, harmonic confinement in radial direction is created by the curvature of the magnetic offset field used to tune interparticle interactions. 
A set of two coil pairs allows for tuning the curvature of the field without changing its offset. 
We can achieve offset fields of up to \SI{1415}{\gauss} and trap frequencies of up to $\omega_r = 2\pi \cdot$ \SI{34}{\hertz}.
This tunability thus allows us either to compensate the optical anti-confinement and create a flat-bottom trap or to provide a harmonic confinement for matter-wave imaging.

\begin{figure*}
\center
\includegraphics[trim=0 2cm 5cm 0cm, clip=true, width = \textwidth]{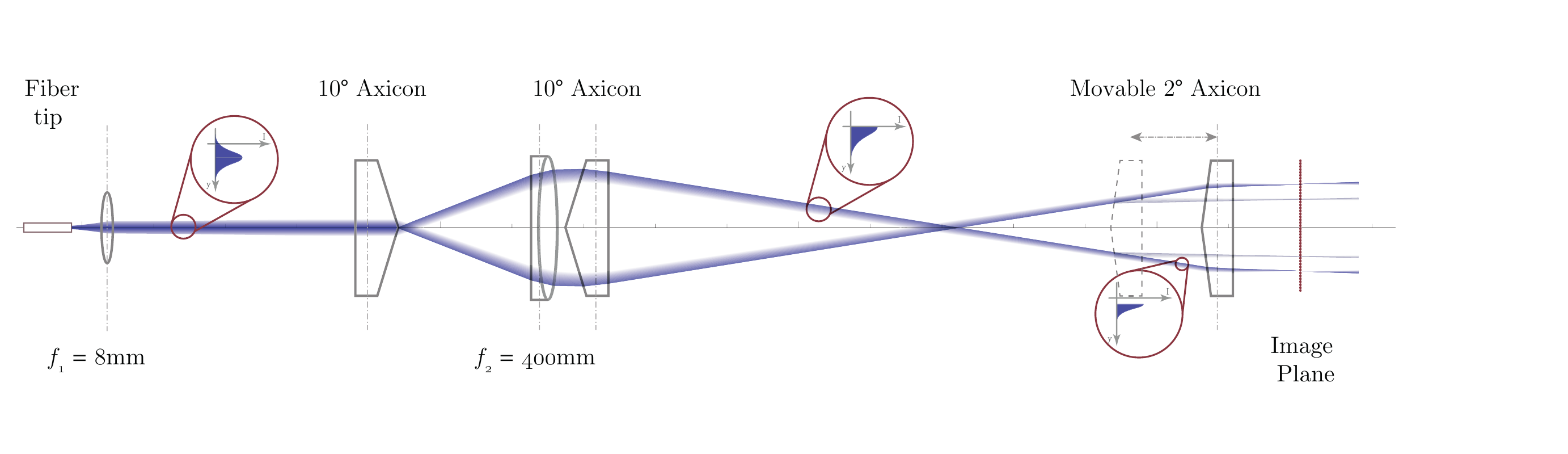}
\caption{\label{figS:Ring}Generation of the ring potential: 
The ring potential is generated by a setup of two lenses and three axicons with different opening angles.
The two lenses image the fiber tip onto an intermediate image plane (red dotted line), which is then imaged onto the atoms using a high resolution objective (not shown).
The first axicon splits the beam into a ring beam which is then optically inverted by the combination of a second axicon and the second lens.
After this optical inversion the steep part of the split Gaussian beam faces towards the center of the ring.
The movable third axicon collimates the geometry of the ring with a variable diameter. 
}
\end{figure*}

\begin{figure}
\center
\includegraphics[width = \linewidth]{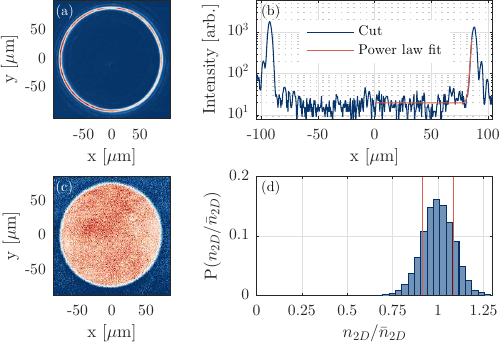}
\caption{\label{figS:DensityDistribution}Characterizing the box potential: 
A cut (b) through an image of the ring potential (a) reveals the steepness of the potential wall. The potential can be approximated by a power law fit $V(x) \propto x^{87\pm4}$. 
Panel (c) shows an averaged density distribution in the box potential at a magnetic offset field of \SI{1100}{\gauss}. 
The flatness of the box potential is quantified by calculating the probability distribution $\text{P}(n_{\text{2D}}/\bar{n}_{\text{2D}})$ for each density $n_{\text{2D}}$ to occur (d). 
The result is well described by a normal distribution with a standard deviation of $\sigma = 0.086\ \bar{n}_{\text{2D}}$ (red solid lines).}
\end{figure}

\section{The Box Potential}

The repulsive optical ring potential is generated by a cascaded setup of two lenses and three axicons (Fig.~\ref{figS:Ring}), which allows tuning the geometry of the ring independently from the beam focus.
The two lenses focus the beam onto an intermediate image plane, which is then imaged onto the atoms using a high resolution objective, while the geometry of the ring is defined by the combination of the three axicons and the second lens. 
The first axicon splits the slightly divergent Gaussian beam into a Bessel beam in the near field and a ring beam in the far field \cite{McLeod54}. 
Together with the second lens, the second axicon leads to an optical inversion of the ring beam such that the steep part of the split Gaussian beam faces towards the center of the ring (see insets in Fig.~\ref{figS:Ring}), resulting in a highly non-Gaussian profile.  
This optical inversion ensures that no residual light -- which is usually present due to imperfections of the axicon tip -- remains in the inner part of the ring.
This has the advantage that no aperture stop has to be placed in the beam path, which was required in previous axicon setups used for trapping ultra cold atoms \cite{Manek98,Mukherjee17}.
Finally, the third axicon collimates the geometry of the ring such that it fits within the NA of the imaging system that projects the ring onto the atoms. 
Moving this axicon along the optical axis allows for the size of the ring to be easily changed. 

The resulting intensity distribution at the position of the atoms can be directly imaged with a second microscope objective (Fig.~\ref{figS:DensityDistribution}~(a)).
A cut through a ring with diameter $D = \SI{160}{\micro\meter}$ is shown in Fig.~\ref{figS:DensityDistribution}~(b) and reveals the steepness of the potential wall.
To obtain a measure of the steepness and gain a model for the shape of the intensity distribution, we fit a power law to one half of the cut and obtain a scaling of $V(x) = x^{87\pm4}$. 

The flatness of the inner part of the box potential can be characterized by measuring the variation of the density $n_{\text{2D}}(\vec r)$ around the mean density $\bar{n}_{\text{2D}}$ in the ring. 
To do this, we take a series of images and calculate the probability of occurrence $\text{P}(n_{\text{2D}}(\vec r)/\bar{n}_{\text{2D}})$ for each normalized density $n_{\text{2D}}(\vec r)/\bar{n}_{\text{2D}}$ \cite{Mukherjee17}. 
Since in a thin 2D sample the fluctuations caused by a corrugated potential can be masked by the inherent quantum fluctuations of the atomic density as well as the photonic shot noise of the imaging, averaging of a sufficient number of images is required.
For our measurement we take an average of 75 density images with constant density, for which the signal to noise ratio (SNR) due to photon shot noise is approximately 29. 
While this makes shot noise negligible, artifacts caused by imperfections of the imaging beam still contribute to the measured fluctuations.
The averaged density distribution together with the corresponding probability distribution $\text{P}(n_{\text{2D}}(\vec r)/\bar{n}_{\text{2D}})$ is shown in Fig.~\ref{figS:DensityDistribution}.
We obtain a standard deviation of the probability distribution of \SI{8.6}{\percent}.

\section{Single/Double Layer Loading and Detection}
\begin{figure}
	\center
	\includegraphics[trim=1.1cm 0 0.5cm 1.1cm, clip=true, width=\linewidth]{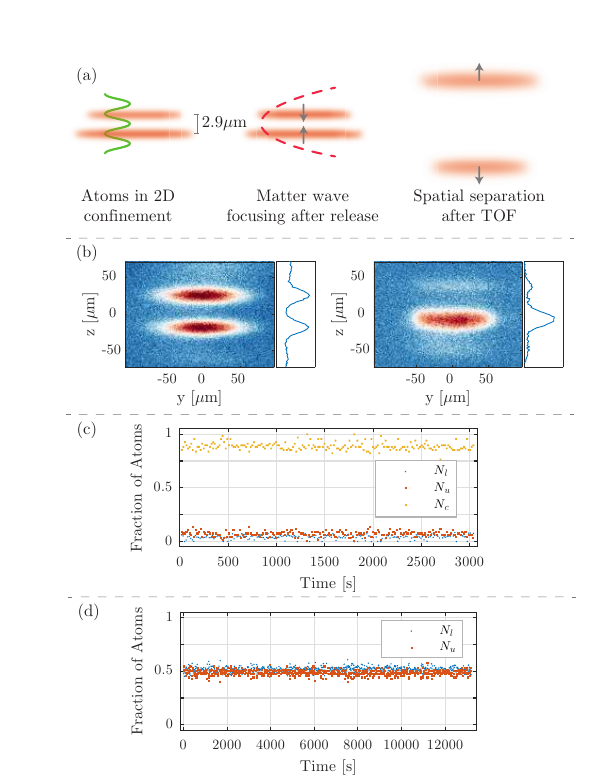}
	\caption{\label{figS:Singlicity}Single/Double Layer Loading: 
	Atoms in two adjacent layers are spatially separated by flashing on an attractive potential in $z$-direction (red dashed line) followed by a time of flight (a). 
	The occupation of individual layers can then be determined by absorption imaging (b).
	Panels (c,d) show measurements of the fraction of atoms in the upper, lower and central layer ($N_u$,$N_l$ and $N_c$) as a function of time for single and double layer loading.}
\end{figure}

To verify the single and double layer loading into the lattice we use a matter wave focusing technique \cite{Murthy14}.
We suddenly switch off the lattice and after a short free expansion, the elliptic trap is flashed on for \SI{18}{\micro\second} followed by a time of flight of \SI{1}{\milli\second}. 
The matter wave lens created by the pulse slows the $z$-expansion of the cloud and at the same time accelerates the different layers with respect to each other (Fig.~\ref{figS:Singlicity} (a)).
This leads to a separation of atoms in different layers after time of flight as shown in Fig.~\ref{figS:Singlicity} (b).
This provides us with a single-shot measurement of the occupation of individual layers, which is a great advantage over other techniques such as RF tomography.
Tracking the occupation of individual layers over time shows that our loading scheme has low fluctuations and is stable on a timescale of several hours (Fig.~\ref{figS:Singlicity} (c,d)).

\section{Density Determination}
The atom density $n_{\text{2D}}(\vec r)$ is determined via high intensity absorption imaging \cite{Hueck171}, where we take into account three additional corrections to the modified Lambert-Beer law.
We correct for the scattering of photons into the NA of the imaging system, pumping of atoms into different hyperfine states and the reduction of the absorption signal when imaging Feshbach molecules.

\begin{figure}
	\center
	\includegraphics[width = \linewidth]{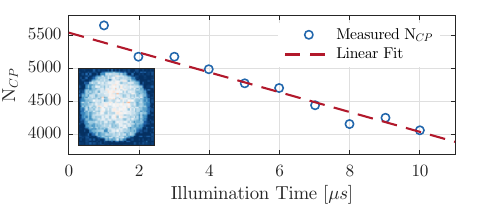}
	\caption{\label{figS:Noft} Detected atom number $N_{CP}$ in the central patch of the cloud in absorption images (inset) as a function of illumination time.
	The apparent atom number drops with longer illumination times as the imaging transition is not fully closed and therefore the number of atoms taking part in the scattering decreases with time (blue open circles).
	A linear fit (red dashed line) results in a correction factor for deducing the true atom number.}
\end{figure}

Absorption imaging is based on the assumption that photons scattered by an atom will not be captured by the imaging system.
However, in systems with a high numerical aperture, a significant fraction of the scattered photons can be recaptured by the imaging system \cite{Pappa11}. 
The correction is given by $\Omega = 2\pi[1-\cos(\theta)]/4\pi$ where $\theta = \sin^{-1}(\text{NA})$ is the opening angle defined by the NA of the imaging system.
For our measurements we have chosen numerical apertures of up to $\mathrm{NA} = 0.4$, which corresponds to a correction of \SI{4}{\percent}.

The second effect we take into account is that the $^2$S$_{1/2}$, F=1/2 to $^2$P$_{3/2}$, F=3/2 optical transition used for imaging the atoms is not fully closed. 
This effect is small for magnetic fields above $\SI{800}{\gauss}$, but becomes significant at lower magnetic fields. 
For our measurements of non-interacting Fermi gases at $\SI{527}{\gauss}$, we therefore take a reference measurement for different imaging durations (see Fig.~\ref{figS:Noft}) and correct the measured density accordingly.

Finally, we take into account that on the bosonic side of a Feshbach resonance, the atoms form weakly bound molecules whose binding energy increases for lower magnetic fields. 
As the molecules become more deeply bound, the atom-light scattering of the bound atoms is altered compared to the behavior of free atoms. 
We correct for this by introducing a reduced scattering cross section $\sigma^\star_0(B)$ for magnetic fields ranging from \SIrange{680}{832}{\gauss}.
The correction is determined by imaging samples with constant atom number at different magnetic fields.

To relate the optical density measured via absorption imaging to the atomic density $n_{\text{2D}}(\vec r)$ we need to know the magnification of our imaging system. 
To measure the magnification we perform Kapitza-Dirac scattering on a lattice potential, which we generate by retro-reflecting the elliptic trap.
This imparts a well-known momentum to the atoms, which allows us to calibrate the magnification by taking images after different times of flight. 
We deduce a magnification of \num{30.8\pm0.3}.

\begin{figure}
\center
\includegraphics[clip=true, width = \linewidth]{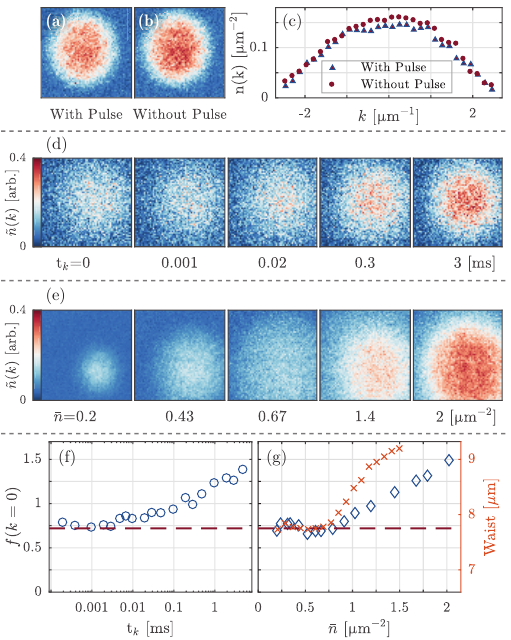}
\caption{\label{figS:FigS5}Influence of collisions and transverse excited states on matter wave focusing. 
To prevent collisions during matter wave focusing, we remove state $\left|\uparrow\right>$ using a resonant light pulse.
To verify that this pulse has only negligible effects on atoms in state $\left|\downarrow\right>$, we compare measurements with (a) and without (b) the pulse in a non-interacting system. 
Line cuts (c) through these images show that the momentum distribution of state $\left|\downarrow\right>$ in a non-interacting system is not significantly altered by the optical removal pulse.
For an interacting system, we map out the effect of collisions on the measured momentum distribution by waiting for a variable interaction time $t_k$ between switching off the optical confinement and removing state $\left|\uparrow\right>$.
With increasing $t_k$, we observe a redistribution of momentum (d,f).
For comparison, the central occupation obtained with $t_k=0$ in Fig.~\ref{fig:Fig4} (c) is marked by red dashed lines.
An additional source for error is that for high 2D densities, low momentum modes in higher transverse levels become occupied.
We record the momentum distribution for different in situ densities $\bar{n}$ (e) and observe an increase in central momentum occupation (blue diamonds, (g)) above a density of $\bar{n} \approx \SI{0.75}{\per\square\micro\meter}$. 
At the same density, we observe an increase in the waist of the cloud after short time of flight (red crosses), which indicates an occupation of higher transverse states.}
\end{figure}

\section{Matter Wave Focusing of interacting 2D Fermi Gases}
When measuring the momentum distribution of interacting gases using matter wave focusing, care has to be taken to ensure that the influence of interactions during the time evolution is negligible. 
While for a 2D gas the density and thus the scattering rate can quickly be reduced by releasing the atoms from the strongly confining potential, the remaining collisions can nevertheless significantly affect the momentum distribution.
To eliminate the influence of collisions, we remove one of the spin components, thereby effectively projecting the system onto a free Fermi gas. 
We accomplish this by illuminating the atoms in state $\left|\uparrow\right>$ with a resonant $\SI{2}{\micro\second}$ light pulse with an intensity of $I=I_{\textit{sat}}$ before performing matter wave focusing. 
This removes the atoms in state $\left|\uparrow\right>$ with a $1/e$ time constant of $\tau_{\uparrow} \approx \SI{150}{\nano\second}$, whereas state $\left|\downarrow\right>$ has a much longer lifetime of $\tau_{\downarrow} \approx \SI{70}{\micro\second}$.
Using a non-interacting Fermi gas, we verify that the momentum distribution of state $\left|\downarrow\right>$ is not significantly altered by the light pulse (Fig.~\ref{figS:FigS5} (a-c)).

We then apply this method to an interacting system at $B=\SI{1020}{\gauss}$.
We map out the influence of interactions on the momentum distribution by switching off the $z$-confinement and removing state $\left|\uparrow\right>$ after different expansion times $t_k$.
We observe that when increasing the time $t_k$ during which collisions can take place beyond $\SI{3}{\micro\second}$, the apparent occupation of lower momentum modes increases (see Fig.~\ref{figS:FigS5} (d,f)).
We attribute this to collisions transferring momentum from the radial into the transverse direction.

We note that the influence of collisions on the matter wave imaged density distribution is small, as the short time during which collisions take place leaves the in situ density distribution largely unaffected. 
Therefore, matter wave images are not a good measure for the influence of collisions on the momentum distribution. 

Finally, the dimensionality of the system has a profound impact on the momentum distribution. 
For low densities, our system is in the 2D regime and we find that $f(k=0)$ remains constant as we increase the density, while the width of the distribution grows. 
As the density surpasses a value of $\bar{n} \approx \SI{0.75}{\per\square\micro\meter}$, $f(k=0)$ begins to increase which we attribute to the population of higher transverse states (see Fig.~\ref{figS:FigS5} (e,g)).
This interpretation is supported by a measurement of the transverse width of the cloud after short time of flight similar to the one described in \cite{Dyke16}.
We find that the transverse width starts to increase at the same density $\bar{n} \approx \SI{0.75}{\per\square\micro\meter}$, signaling the beginning of the crossover to a 3D system. 
\end{document}